%% file: main.tex
\documentclass[letterpaper,twocolumn,10pt]{article}
\usepackage{usenix-2020-09}
\usepackage{xurl}
\usepackage{graphicx}
\usepackage{booktabs}
\usepackage{longtable}
\usepackage{xspace}
\usepackage{listings}
\usepackage{pgfplots}
\pgfplotsset{compat=1.18}
\usetikzlibrary{positioning}
\pgfplotsset{discard if not/.style 2 args={x filter/.append code={%
  \edef\tempa{\thisrow{#1}}\edef\tempb{#2}%
  \ifx\tempa\tempb\else\fi}, unbounded coords=discard}}
\definecolor{closedc}{HTML}{2A78D6}
\definecolor{openc}{HTML}{EB6834}
\lstdefinelanguage{Rust}{
  morekeywords={fn,let,mut,use,pub,impl,struct,unsafe,move},
  sensitive=true, morecomment=[l]{//}, morestring=[b]"}
\input{metrics}
\newcommand{\sys}{\textup{\textsc{Crustify}}\xspace}
\newcommand{\req}[1]{\texttt{#1}}
\newsavebox{\takeawaybox}
\newenvironment{takeaway}{\par\medskip\noindent\begin{lrbox}{\takeawaybox}%
  \begin{minipage}{\dimexpr\linewidth-2\fboxsep-2\fboxrule}\itshape}{\end{minipage}%
  \end{lrbox}\fbox{\usebox{\takeawaybox}}\par\medskip}

\title{Agentic Discovery of Soundness Bugs in Safe Rust Bindings}
\author{
{\rm Marius Momeu}\\
University of California, Berkeley\\
\texttt{marius.momeu@berkeley.edu}
\and
{\rm Koushik Sen}\\
University of California, Berkeley\\
\texttt{ksen@berkeley.edu}
}

\begin{document}

\maketitle

\input{sections/abstract}

\input{sections/introduction}
\input{sections/background}
\input{sections/study}
\input{sections/evaluation}
\input{sections/design}
\input{sections/wild}
\input{sections/case-studies}

\input{sections/limitations}
\input{sections/related}
\input{sections/conclusion}

\bibliographystyle{plain}
\bibliography{references}

\appendix
\input{sections/appendix}

\end{document}

%% file: metrics.tex
\newcommand{\numCrates}{19}
\newcommand{\numBindings}{16}

\newcommand{\numConsumers}{three}
\newcommand{\numConsumerOwn}{4}
\newcommand{\numConsumerReached}{5}
\newcommand{\numAudits}{22}
\newcommand{\numFindings}{269}

\newcommand{\numReported}{78}
\newcommand{\numFixed}{24}
\newcommand{\numFixedOthers}{25}
\newcommand{\numToFile}{118}
\newcommand{\numDoNotFile}{48}
\newcommand{\numKnownUpstream}{24}
\newcommand{\numNits}{24}
\newcommand{\costLow}{3}
\newcommand{\costHigh}{251}
\newcommand{\costTotal}{1{,}349}

\newcommand{\costPerFinding}{5.17}

\newcommand{\costMedian}{42}
\newcommand{\costMedianCeil}{50}
\newcommand{\costSecond}{218}
\newcommand{\findingsPerCrateLow}{3}
\newcommand{\findingsPerCrateHigh}{59}

\newcommand{\numUB}{213}
\newcommand{\numUBUnsafe}{22}
\newcommand{\numEquiv}{23}
\newcommand{\numContracts}{33}
\newcommand{\numContractsNits}{22}

\newcommand{\numReportedPublic}{76}

\newcommand{\numIssues}{73}

\newcommand{\studyRuns}{24}
\newcommand{\studyArms}{8}
\newcommand{\studyFindings}{838}
\newcommand{\studyFalse}{one}

\newcommand{\studyMeetsWithoutLow}{17}
\newcommand{\studyMeetsWithoutHigh}{51}
\newcommand{\studyMeetsWithLow}{74}
\newcommand{\studyMeetsWithHigh}{81}
\newcommand{\studyNonPublic}{51}
\newcommand{\studyRealDistinct}{86}

\newcommand{\studyFullOpen}{25}
\newcommand{\studyFullClosed}{26}
\newcommand{\studyFullBoth}{18}
\newcommand{\studyOnce}{15}
\newcommand{\studyHalf}{14}
\newcommand{\studyCostClosed}{0.67}
\newcommand{\studyCostOpen}{0.17}
\newcommand{\studyCost}{87}

\newcommand{\veightRepro}{52}
\newcommand{\veightRerunSame}{49}
\newcommand{\rflFindings}{15}
\newcommand{\rflUB}{11}
\newcommand{\rflContracts}{4}
\newcommand{\rflKASAN}{8}
\newcommand{\rflWall}{1\,h\,45\,min}

\newcommand{\studySanCauses}{35}
\newcommand{\studySanInstances}{505}
\newcommand{\studySanInstancesSan}{173}
\newcommand{\studySanInstancesNone}{231}

\newcommand{\manualN}{50}

\newcommand{\rerunRunnable}{33}
\newcommand{\rerunRan}{33}
\newcommand{\rerunReproduced}{31}
\newcommand{\rerunPartial}{one}
\newcommand{\rerunNot}{one}
\newcommand{\rerunNone}{17}

\newcommand{\suppRunnable}{15}
\newcommand{\suppContractsArgued}{9}
\newcommand{\suppContractsShown}{6}
\newcommand{\secondAgree}{47}
\newcommand{\secondContractsReal}{14}

%% file: sections/abstract.tex
\begin{abstract}
Rust crates confine \texttt{unsafe} calls into C and C++ code behind safe
interfaces that must uphold Rust's invariants. If an interface encodes a rule of the code beneath it too weakly,
a program in purely safe Rust can cause undefined behavior. Static
analyzers, bound to fixed patterns, and fuzzers, bound to the paths they
execute, often miss such bugs. We propose LLM agents as auditors of these interfaces,
and study how well they do it.

We classify these defects and measure, in repeated audits of two
crates by a closed and an open model, Claude Opus~5.5 and DeepSeek V4.1 Flash,
what agents find, how well they triage, and what makes them find more. Agents
find undefined behavior and divergences from the wrapped library unprompted,
but look at \texttt{unsafe} contracts only when asked; their findings are
almost always real, yet they seldom confirm them with a sanitizer and stop
early. With the same guidance, the open model finds as many defects as the
closed one at a quarter of the cost, and where few defects are left, a target
for the number of findings that cannot be met makes neither fabricate findings.

We build these insights into the audit agent of \sys, an agentic system for
C-to-Rust tasks, and run it on \numBindings{} widely used bindings to C and C++
libraries, such as OpenSSL, SQLite, libgit2, and V8, on \numConsumers{} crates
that use them, such as \texttt{cargo}, and on the Rust abstractions of the
Linux kernel.
It found \numUB{} undefined-behavior bugs, \numContracts{} incomplete contracts,
and \numEquiv{} divergences in the crates, at a median cost under
\$\costMedianCeil{} per audit, and \rflUB{} undefined-behavior bugs in the kernel.
Maintainers have fixed \numFixed{} of the \numReported{} findings that we reported.
\end{abstract}

%% file: sections/introduction.tex
\section{Introduction}
\label{sec:intro}

Rust guarantees memory and thread safety for programs that contain no
\texttt{unsafe} code. Much of its ecosystem nevertheless rests on code that the
compiler cannot check. Crates that wrap C and C++ libraries, such as OpenSSL,
SQLite, and RocksDB, or the operating system's interface, such as \texttt{nix} and
\texttt{socket2}, call foreign functions through the foreign function interface (FFI) in
\texttt{unsafe} blocks and expose
them through an API that is marked safe~\cite{astrauskas2020unsafe}. Such a
binding is \emph{sound} if no program written in purely safe Rust can cause undefined behavior (UB) through it~\cite{jung2018rustbelt}. Since applications call these APIs on untrusted input, an unsound binding can turn a program written in safe Rust into one with a memory-safety vulnerability, and a single unsound
function voids Rust's guarantees for every program that uses it. The same
holds for the safe abstractions on which Rust drivers in the Linux kernel are
built.

The binding's authors must encode in Rust's types the rules that the wrapped library
states in its documentation, or only in its implementation: which call frees
an object, which pointer a later call invalidates, which handle may be shared
between threads. Listing~\ref{lst:motivating} shows how easily this goes
wrong: \texttt{Session::auth\_methods} of the \texttt{ssh2} crate returns a
string that points into a buffer that libssh2 frees on the next call, and the
borrow checker permits that call, so the program reads freed memory without a
single \texttt{unsafe} block.

\begin{figure}[t]
\begin{lstlisting}[language=Rust, label={lst:motivating},
  caption={A use-after-free in safe Rust through \texttt{ssh2}~0.9.7 (issue~\#378). The second call frees the buffer that \texttt{first} points into.}]
#![forbid(unsafe_code)]
fn main() {
    let mut sess = ssh2::Session::new().unwrap();
    sess.set_tcp_stream(TcpStream::connect(ADDR).unwrap());
    sess.handshake().unwrap();
    let first: &str = sess.auth_methods("root").unwrap();
    let _second = sess.auth_methods("root").unwrap();
    println!("{first}"); // reads freed memory
}
\end{lstlisting}
\end{figure}

Existing techniques are ill-suited for finding such bugs. Static analyzers
detect instances of the patterns that they
encode~\cite{bae2021rudra,nitin2024yuga}, rarely model the foreign
library~\cite{li2022ffichecker}, and report candidates rather than bugs.
Sanitizers~\cite{serebryany2012asan,serebryany2011tsanllvm} and
Miri~\cite{miri2026} report a violation only when an execution exhibits it, and
a crate's tests exercise its API as its authors intended. Fuzzers for Rust
libraries generate sequences of calls from API
signatures~\cite{jiang2021rulf,deepsurf2025}, but leave out crates that
depend on a foreign library, and a soundness bug often depends on a rule of
that library that no signature states. These bugs require an auditor that
reads the Rust code together with the foreign code that it calls, infers the rule
that the binding encodes too weakly, and writes the program that breaks it;
the instruments then confirm that the program has undefined behavior.

LLM agents have recently become good at this kind of work: they read code,
write programs, and run them. Claude Mythos Preview, run in Claude Code
with a one-paragraph prompt, found thousands of high-severity vulnerabilities,
among them bugs in every major operating system and web browser, and wrote
working exploits for many~\cite{anthropic2026mythos}; Revelio has agents propose memory-safety bugs
in C and C++ and confirms each with a sanitizer~\cite{hou2026revelio}. These
bugs show up in an execution of the existing program, whereas a soundness bug
shows up only in a program that the auditor must write. Prior work on auditing
bindings pairs the model with a fuzzer that finds the
bug~\cite{deepsurf2025}, or has it return a verdict on contracts rather than a
demonstration~\cite{li2025safe4u}, and practitioners report that an agent
alone finds soundness bugs~\cite{davidoff2026unreasonable,microsoft2026forge}.
How good agents are at this task, what they miss, and what an agent harness
for auditing needs has not been measured.

Measuring this poses four problems. First, audits are
not deterministic: two audits of the same code with the same prompt find
different defects. Second, there is no ground truth, so recall cannot be
computed. Third, an agent that reads the network can restate defects from the
project's tracker rather than discover them. Fourth, agents file hundreds of
findings, and each must be judged on whether it is real, whether it is one
defect or several, and whether its evidence shows what it claims.

We address these problems as follows. We name the classes of defects at
this boundary (\S\ref{sec:taxonomy}):
\emph{objectives} that a crate is expected to meet, e.g., freedom from undefined
behavior (\req{ub}), sound contracts of \texttt{unsafe} functions
(\req{contracts}), and equivalence to the wrapped library (\req{equiv}), and nine
\emph{hazards} through which such code reaches undefined behavior. We state
\emph{triage rules} that every finding must meet: one root cause, stored
evidence such as a reproducer that fails under a sanitizer, and no claim
beyond that evidence. In a controlled study (\S\ref{sec:study}), we hold the
harness and the instruments fixed and vary one file of guidance, built up from
minimal guidance, the request alone, to the full prompt. We run every arm on an
open model, DeepSeek V4.1 Flash, and the arms with no guidance and with all of
it also on a closed model, Claude Opus~5.5, and repeat each configuration three
times to account for variance. In place of
recall, we count the distinct defects that are not
public upstream, which an agent cannot have copied from the tracker. A judge
model scores every finding blind to the arm and the model; for a random sample,
a second model scores it again and agents rerun the reproducers. We audit the
\texttt{git2} crate, in which our earlier audits had found many defects; the
audits file \studyFindings{} findings at a cost of \$\studyCost{}. A second
experiment runs the full prompt with both models, again three times each, on
\texttt{aws-lc-rs}, in which our earlier audits had found few, and tests
whether a target that cannot be met makes agents fabricate findings. The study
yields four insights (\S\ref{sec:eval}).

\paragraph{Insight \#1: Agents understand these defects.} The judge rates only \studyFalse{} of
the \studyFindings{} findings false. Of the \studyRealDistinct{} distinct real defects
that the audits found, \studyNonPublic{} are not public upstream, so no agent can
have copied them from the tracker; with no guidance at all, both models
find undefined behavior among them. They identify it correctly
whether or not they run an instrument: of the \studySanInstances{} findings of
defects that a sanitizer confirmed in some audit, \studySanInstancesNone{} store no
instrument's report and rest on the code alone or on the upstream report, and
these are real as well, although some only point to the upstream report. An
auditor can thus find these bugs without a fuzzer or a static analyzer, as practitioners have also
observed~\cite{davidoff2026unreasonable}; the instrument confirms them, so
that a maintainer need not take the agent's word.

\paragraph{Insight \#2: Agents look only where they are pointed.} Unprompted,
agents look for memory-safety bugs and for defects of behavior, and for nothing
else. No unprompted audit of either model finds an insufficient contract of an
\texttt{unsafe} function that is not already public; when \req{contracts} is named, both models look at contracts, but demonstrate few of the defects that they file, in a crate that has few to find. Named, unnecessary \texttt{unsafe} code yields many findings: a single audit filed 28, none of them judged false, each with a patch that removes the \texttt{unsafe} code and passes the crate's tests. An auditor must therefore name every objective for which it wants findings.

\paragraph{Insight \#3: Agents do not triage unless told.} Without the triage
rules, at most \studyMeetsWithoutHigh{}\% of the findings meet them: the others
rest on an argument rather than a run, bundle several root causes, or split one
into several findings. The unprompted closed model stored a sanitizer's report
for three of its 99 findings. With the rules, at least \studyMeetsWithLow{}\% of
the findings meet them, on both models. Agents can produce the evidence, but
do not consider it part of the task.

\paragraph{Insight \#4: With guidance, an open model audits as well as a closed one.}
With all guidance, the three audits of each model find \studyFullOpen{} and
\studyFullClosed{} non-public defects, \studyFullBoth{} of them in common, and
the open model costs a quarter as much per defect. Agents end an audit when
they consider it done, before the budget is spent, the open model
long before; a stated target for the number of findings is the one piece of
guidance that consistently raises how many unknown defects the open model
finds. In a crate with few defects left, the target does not make agents
fabricate findings: under a target that neither can meet, both models file
only real findings, one or two per audit, and say that they fell short. Audits are only partly reproducible: of the
\studyNonPublic{} distinct non-public defects that the study found,
\studyOnce{} were found by a single audit. The union of several cheap audits
therefore finds more than any one of them.

\paragraph{\sys at scale.} We integrate these insights in
\sys, an agentic system for automating C-to-Rust tasks~\cite{momeu2026crustify}, as a new audit agent for
popular open-source crates that call C or C++ code through the FFI (\S\ref{sec:design}): its prompt names the
objectives, states the triage rules, and sets a minimum number of unknown
defects that an audit must find before it may end. With \sys, we audited \numCrates{} widely used crates and
confirmed \numFindings{} distinct findings, \numUB{} of them undefined
behavior, all but nine with a reproducer that independent agents reran, as did we for every finding that we reported. We have
reported \numReported{} of them, and maintainers have fixed \numFixed{},
including use-after-frees in \texttt{openssl}'s certificate verification
and in \texttt{rusqlite}, and a data race and use-after-frees in \texttt{mlua}
(\S\ref{sec:wild}). In the Rust abstractions of the Linux kernel, which no
user-space instrument supports, \sys filed \rflFindings{} findings, \rflKASAN{}
confirmed by the kernel's address sanitizer.

\paragraph{Contributions.} We make the following contributions:
\begin{itemize}
  \item A classification of the defects of Rust code that interfaces with C and C++, by
  objective and by hazard, and triage rules for findings
  (\S\ref{sec:taxonomy}, \S\ref{sec:triage}).
  \item The first systematic study of LLM agents as auditors of such code,
  with repeated audits and blind scoring (\S\ref{sec:study},
  \S\ref{sec:eval}).
  \item An open-source auditor in \sys, built on the study's results, which
  needs neither a fuzzer nor a static analyzer (\S\ref{sec:design}), at
  \url{https://github.com/crustify-rs}.
  \item \numFindings{} findings in \numCrates{} widely used crates
  and \rflFindings{} in the Linux kernel (\S\ref{sec:wild}).
\end{itemize}

%% file: sections/background.tex
\section{Background}
\label{sec:background}

\subsection{Soundness of Safe Bindings}
\label{sec:bg-soundness}

\paragraph{Safe and unsafe Rust.} Every Rust value has a unique owner that
releases it, and other code accesses it through shared references
\texttt{\&T} or one mutable reference \texttt{\&mut T}, whose
\emph{lifetime} the borrow checker bounds by that of its referent. The marker
traits \texttt{Send} and \texttt{Sync} state that a type may be moved to, or
shared with, another thread. Inside
\texttt{unsafe} blocks, a program may also dereference raw pointers, call
\texttt{unsafe} functions, including foreign ones, and implement \texttt{Send}
and \texttt{Sync} by hand; the compiler does not check these operations, and a
program that performs them incorrectly has undefined
behavior~\cite{rustreferenceub}.

\paragraph{Soundness.} Soundness is a property of an API, not of its callers: a
safe function must be safe for every argument that type-checks, in every order
and on every thread that the borrow checker permits. A function that requires
more of its caller, e.g., a valid pointer, must be marked \texttt{unsafe} and
document the conditions under which calling it is sound. An unsound function
is a defect even if no existing program calls it incorrectly.

\paragraph{The FFI boundary.} A binding's \texttt{-sys} crate declares the foreign
library's functions and types, typically generated by bindgen~\cite{bindgen},
and its safe crate wraps them in types that own or borrow the library's
objects~\cite{git2rs,rusqlite,rustopenssl,curlrust}. The wrapper must encode
the library's invariants, e.g., which function releases an object or when a
registered callback runs, which C does not state in its types and a C++ API states only in part, as ownership,
lifetimes, and trait bounds. The Rust abstractions of
the Linux kernel stand in the same relation to the kernel's C interfaces.

\subsection{Detecting Undefined Behavior}
\label{sec:bg-instruments}

Sanitizers instrument a program at compile time and report an execution that
violates the invariants they check: AddressSanitizer
(ASan)~\cite{serebryany2012asan} temporal and spatial memory errors,
UndefinedBehaviorSanitizer (UBSan)~\cite{ubsan} operations that are undefined in
C, ThreadSanitizer (TSan)~\cite{tsan2009,serebryany2011tsanllvm} data races, LeakSanitizer (LSan, part of ASan) memory leaks, and MemorySanitizer (MSan)~\cite{msan2015} reads of uninitialized memory.
Since sanitizers operate on the compiler's intermediate representation, they
instrument Rust and C alike and detect errors that span the two, but they do
not check the rules specific to Rust, such as the validity of values or the
aliasing of references. Valgrind detects memory errors in an uninstrumented
binary. Miri~\cite{miri2026}, an interpreter for Rust's
mid-level intermediate representation, checks these rules, including the
Stacked and Tree Borrows aliasing
models~\cite{jung2020stackedborrows,villani2025treeborrows}, but runs
foreign code only in an experimental mode, which checks none of these rules
inside it. BorrowSanitizer (BSan)~\cite{mccormack2025bsan} instruments Rust
together with statically linked C to detect violations of Tree Borrows,
including those that an access on the C side causes. In the kernel, the Kernel Address Sanitizer (KASAN) and the Kernel Concurrency Sanitizer (KCSAN) take the roles of ASan and TSan, and KUnit tests run in-kernel code under them. We call these tools \emph{instruments}.

All of them report a violation only on the paths that an execution takes:
McCormack et al.~\cite{mccormack2025mirilli} found undefined behavior in dozens
of crates that call foreign functions by running their existing tests under an
extended Miri, but a soundness bug needs a program that misuses the API in a
way that its authors did not anticipate.

\subsection{LLM Agents}
\label{sec:bg-agents}

An LLM agent is a model that calls tools in a loop, e.g., to read a file or run
a command, until it considers its task done~\cite{yao2023react}. A
\emph{harness}, such as Claude Code~\cite{claudecode} or Codex~\cite{codex},
runs this loop and supplies the tools and a \emph{system prompt}, to which a
user or a plugin can append text. An agent can start \emph{subagents} that each
receive a brief, work in their own context, and return a result.

%% file: sections/study.tex
\section{Study Design}
\label{sec:study}

Before we design an auditor, we ask what agents can already do at this task
and what they lack. We answer the following research questions:
\begin{itemize}
  \item \textbf{RQ1 (Understanding).} Do agents find real defects in Rust code
  that interfaces with C, and how many of their findings are false?
  \item \textbf{RQ2 (Coverage of the objectives).} Which classes of defects do
  agents find unprompted, and which only when the class is named?
  \item \textbf{RQ3 (Triage).} How many findings meet the triage rules, and what
  does stating the rules change?
  \item \textbf{RQ4 (Guidance and discovery).} Which part of the guidance raises
  the number of unknown defects that an audit finds?
  \item \textbf{RQ5 (Models, cost, and reproducibility).} How do a closed and an
  open model compare, at what cost, and how reproducible is an audit across
  repeated runs?
  \item \textbf{RQ6 (In the wild).} What do such audits find in widely used
  crates, and how do maintainers respond?
\end{itemize}
A controlled study on one crate, with a second for the target (\S\ref{sec:study-target}), answers
RQ1 to RQ5 (this section and \S\ref{sec:eval}); we integrate its results in \sys (\S\ref{sec:design}) and answer
RQ6 with audits at scale (\S\ref{sec:wild}).

\subsection{The Auditing Agent}
\label{sec:overview}

The agent that we study is given a Rust crate at a revision, and optionally
the objectives in scope, the build configuration, and a budget of dollars and
time. It produces \emph{findings}: one directory per defect, holding a report that could be sent to the project that owns the defect, and the evidence. A \emph{defect} is a place where the crate falls short of an objective (\S\ref{sec:tax-objectives}), and its \emph{class} is that objective; a defect of \req{ub} is a \emph{bug}. We count defects by \emph{root cause}, what one coherent fix removes: a finding may bundle several root causes, and several findings may repeat one.

The agent works in a container that holds the crate's checkout and the
toolchains for every instrument of \S\ref{sec:bg-instruments}, applied to the
Rust code and the C library alike. It can read the network, e.g., the issue
tracker and the C library's documentation, but sends nothing: a person decides
what is reported. The agent decides by itself what to read, in what order, and
whether to divide the work among subagents. No fuzzer and no static analyzer
takes part: the agent finds a defect by reading the Rust code and the C code
that it calls, and shows it by writing a program and running it.

\subsection{Guidance}
\label{sec:guidance}

An auditor can give the agent guidance as text that precedes the request. We
study four pieces of guidance, each a section of \sys's audit prompt: the
objectives that a crate must meet, the hazards through which it falls short of
the first, the rules by which leads become findings, and a target for their
number. The objectives and the hazards also name what there is to find: we
classify every finding by them.

\subsubsection{Objectives}
\label{sec:taxonomy}
\label{sec:tax-objectives}

An \emph{objective} is a property that a crate that interfaces with FFI code is
expected to meet and on which its consumers rely. We study four, which concern
the safety and the correctness of the boundary.

\paragraph{Free of undefined behavior (\req{ub}).} No use of the API that
meets its documented contracts can cause UB: this is soundness as
\S\ref{sec:bg-soundness} defines it. A defect is shown by a program, in
purely safe Rust or meeting every documented contract that it relies on, that
causes UB under an instrument. In the latter case the contract is defective too; we
count such a finding once, where its audit filed it.

\paragraph{Sound contracts (\req{contracts}).} An \texttt{unsafe} function is
exempt from soundness; instead, its documentation must state the conditions
under which calling it is sound, and every \texttt{unsafe} block must say why
its conditions hold. A contract that omits a condition, or states
one that does not suffice, leaves a caller who follows it with UB.

\paragraph{Equivalent to the library (\req{equiv}).} A caller of the binding
observes what a caller of the wrapped library observes: the same results,
errors, output, and side effects. A divergence, e.g., an error that the
wrapper drops or an option that it maps to the wrong constant, is a defect of
behavior rather than of memory safety, and we count such findings separately
throughout.

\paragraph{Minimal \texttt{unsafe} code (\req{unsafety}).} No \texttt{unsafe}
code exists where safe code can do the same work. Unnecessary \texttt{unsafe}
code is not a bug while it is sound, but it is a risk: the compiler does not
check it, every audit must trust it, and a later change can break the invariant
it relies on.

The controlled study evaluates the first three objectives, and a further
experiment the fourth (\S\ref{sec:further}). \sys's prompt also names
objectives of software engineering, e.g., the usability and the stability of
the API, idiom, tests, and performance; this paper does not evaluate them.

The \emph{Objectives} section of the prompt lists every objective with its
identifier, and tells the agent to file each finding under the objective that
the crate does not meet and to audit only the objectives that the user names.

\subsubsection{Hazards}
\label{sec:tax-hazards}

The bugs under \req{ub} share the structure of Listing~\ref{lst:motivating}: a
safe interface encodes a rule of the \emph{code beneath} it too weakly, and a
program that breaks no documented contract violates the rule. The code beneath is
whatever the interface hides from the compiler: a C or C++ library, the operating
system's interface, the kernel's C, or the crate's own \texttt{unsafe} Rust.
Table~\ref{tab:hazard-defs} distinguishes nine hazards by the rule that is
violated. Each holds in both directions: when Rust calls into the code beneath,
and when that code calls back into Rust or keeps pointers into Rust memory. The
list is a starting point rather than a partition: a defect that fits none of
the hazards is still a defect. We use the hazards chiefly to classify findings
(Table~\ref{tab:hazards}); as guidance, they do not make agents find more
(\S\ref{sec:rq4}). Aliasing violations do not corrupt memory with current
compilers, and Rust has not yet adopted an aliasing model normatively; we count
them as undefined behavior, as Miri does, and list them separately. The
\emph{Hazards} section of the prompt lists them as a starting point for the
search, phrased for a binding to a C library (Appendix~\ref{sec:prompts}).

\begin{table}[t]
\centering
\caption{The hazards of undefined behavior, by the rule that is violated.
We classify every instance of undefined behavior that the audits find by these
hazards (Table~\ref{tab:hazards}).}
\label{tab:hazard-defs}
\small
\begin{tabular}{@{}p{0.24\linewidth}p{0.70\linewidth}@{}}
\toprule
Hazard & The rule that is violated \\
\midrule
Lifetimes & A reference, slice, or handle outlives the memory it refers to,
which the code beneath frees or reuses on a later call; or the code beneath
keeps a pointer to a Rust value that moves or is dropped. \\
Callbacks, re-entrancy & The code beneath calls into Rust after the closure or
its data is gone, from within an operation that the callback re-enters, or on
another thread; or a panic unwinds across the boundary. \\
Threads & \texttt{Send} or \texttt{Sync} promises what the wrapped object does
not have, or state that must be locked is reached without the lock. \\
Aliasing & A mutable access derives from a shared borrow, or the code beneath
writes behind a shared reference (a Stacked or Tree Borrows violation). \\
Untrusted values & A length, index, pointer, or return value crosses the
interface unchecked; data is not a valid value of its type, or is read
uninitialized. \\
Ownership & An object is freed twice, by the wrong side, or with the wrong
deallocator; a once-only operation runs twice; an invariant relies on a
destructor that safe code may skip. \\
Integers & A conversion between the two sides truncates or changes sign, or
an unchecked value makes the code beneath overflow. \\
Layout & A Rust type differs from the type on the other side, an access is
misaligned or too large, or an opaque type is used as a value. \\
Unsafe APIs & A caller meets every documented condition of an
\texttt{unsafe} function and still causes UB, so the contract is the defect. \\
\bottomrule
\end{tabular}
\end{table}

\subsubsection{Triage Rules}
\label{sec:triage}

A \emph{lead} is a suspected defect. The goal of triage is that every real
defect among the leads becomes a finding that a maintainer can accept without
trusting the agent, and that nothing else does. The \emph{Triaging} section
states this goal as rules:
\begin{itemize}
  \item \emph{Decide every lead.} A lead ends as a finding or as not a defect,
  and is dropped only on evidence against the defect itself, not on a wrong
  explanation or a trigger that has not been found yet.
  \item \emph{One finding per root cause.} A finding names the one fix that
  removes it; several root causes are not put in one finding, and one is not
  filed several times.
  \item \emph{Evidence.} Every finding carries the strongest kind of evidence
  that the defect allows (Table~\ref{tab:evidence}).
  \item \emph{Claim only what the evidence shows.} The objective, the reach, and
  the severity that a finding states are limited to what the stored output of
  its evidence shows.
\end{itemize}

\begin{table}[t]
\centering
\caption{The three kinds of evidence, from strongest to weakest. A finding
carries the strongest kind that its defect allows.}
\label{tab:evidence}
\small
\begin{tabular}{@{}lp{0.72\linewidth}@{}}
\toprule
Kind & What is stored with the finding \\
\midrule
Reproducer & Where the defect can be triggered: a program that triggers it
through the affected API, the command that runs it, and its output, e.g., the
report of a sanitizer or of Miri, or an assertion against the behavior of the C
library \\
Witness & Where the defect cannot be triggered but can be measured: a change or
a measurement that shows the shortfall, with its output \\
Review & Where the defect is a matter of judgment: the record of an agent that
did not find the lead, tried to refute it, and failed \\
\bottomrule
\end{tabular}
\end{table}

A reproducer must not do what the finding says the code allows: if the finding
is about the safe API, the reproducer contains no \texttt{unsafe} code, which
\texttt{\#![forbid(unsafe\_code)]} lets the compiler check; if it is about an
\texttt{unsafe} function, it meets every obligation that the function
documents. The rules are also the bar against which we score every finding,
including those of agents never shown them (\S\ref{sec:study-scoring}).
\sys's prompt has three further rules about known defects (search upstream
before reproducing a lead, end a lead whose root cause is already reported, and
write no reproducer for it), which the study removes from every arm
(\S\ref{sec:study-arms}).

\subsubsection{A Target}
\label{sec:target}

The \emph{Fitness} section of the prompt states that an audit has two goals
within the user's budget: recall, i.e., to find as many of the crate's defects
as possible, and precision, i.e., to file only findings that meet the triage
rules. An audit is fitter when it has more findings and every one meets the
rules; a finding that does not meet them lowers the fitness. The section ends
with a default target: the audit is not done before it has 25 findings that
meet the rules, unless the user asks for more or fewer, and triaged leads and
reviewed files are not reasons to stop.

\subsection{Subjects, Models, and Arms}
\label{sec:study-target}
\label{sec:study-models}
\label{sec:study-arms}

The controlled study audits \texttt{git2}~\cite{git2rs}, the binding to libgit2,
at revision \texttt{b9d59b6}: the safe crate, built with one fixed set of
features; the \texttt{-sys} crate and the vendored C library are out of scope
except as they explain a defect in the crate. Since one audit per
configuration says little, every arm runs three times, and we report each
audit, what the audits of an arm have in common, and what only one finds. We
sort every real defect into \emph{public}, if an issue or a pull request of the
project describes its root cause, and \emph{not public} otherwise, and count
discovery on the second group, since a public defect can be copied from the
tracker. Some public issues are our own reports from earlier audits of this
crate (\S\ref{sec:wild}). Pilot rounds, run while we developed the method, are
not reported here. We chose \texttt{git2} because our earlier audits had found
many defects in it, so that an audit has much to find and the arms can differ.

A target for the number of findings could make an agent fabricate findings
once real ones run out. To test this, we audit a crate in which our earlier
audits had found few defects: \texttt{aws-lc-rs}~1.18.2 (\texttt{a5ec447}),
the binding to AWS-LC, in which they had found three instances of undefined
behavior and mostly nits otherwise. Each audit receives \sys's full prompt,
whose target asks for 25 unknown defects, and a request for \req{ub} alone,
with the default backend and two further features. Each model ran three
times, the closed model with \$50 and four hours, and the open model with \$10
and four hours. The same judge scored the six audits blind.

We study one closed and one open model: Claude Opus~5.5 and DeepSeek V4.1 Flash
(\texttt{deepseek-v4.1-flash-1017}). Both run in the same harness, Claude
Code~\cite{claudecode}, in the same container image, and are served through
OpenRouter with one pinned provider per model. A proxy meters every request,
including those of subagents, and stops the audit when its budget of dollars
or of time is spent. The container holds no credentials, so that an agent can
read the network but cannot post to it.

The arms differ in one file, which the harness appends to its own system
prompt. The \emph{none} arm
appends nothing; the \emph{+objectives}, \emph{+hazards}, \emph{+triage}, and
\emph{+target} arms append one section of the guidance each
(\S\ref{sec:guidance}, \S\ref{sec:triage}); and the \emph{full} arm appends
\sys's audit prompt with all four, which also holds a short preamble, a
suggestion to divide the work among subagents, and a rule to work without
asking when the user cannot be asked. The open model runs all six arms, and the
closed model \emph{none} and \emph{full}, each three times.

All arms receive the same request: the crate, its revision, the scope, the
build configuration, where to put the findings, the budget, and the instruction
to work without asking and to post nothing. The arms whose file contains the
objectives are asked to audit against \req{ub}, \req{contracts}, and
\req{equiv} and no other objective; the others are asked to audit the crate
``for defects''. The budget is \$10 and three hours per audit for the open
model, and \$15 and three hours for the closed one
(Appendix~\ref{sec:prompts}). We do not ask any arm to leave known defects out,
since an agent could then meet the request by searching the tracker; every arm
is asked to report every defect that it finds and to link the upstream report
if one exists. To rule out leaks, the directory that would hold the plugin is
empty in every arm, and after every audit we scan its transcript for the paths
of the prompt files and for marker phrases of the sections that the arm did not
receive.

\subsection{Scoring}
\label{sec:study-scoring}

Scorers see each audit's output under a random code, without the arm or the
model, and judge every finding by reading the code, the report, and the stored
evidence: the root causes that it covers, whether each is real, the kind of
evidence stored, whether it claims more than its evidence shows, and whether
it repeats a root cause that the audit already filed. The scorer is a frontier
model, Claude Fable~5.2, used as a judge~\cite{zheng2023judge}: each instance
receives a few audits under their codes, the crate's checkout, and our
instructions (Appendix~\ref{sec:prompts}), but not the arms, the models, or the
other scores. We merged by hand the names that scorers gave to the same new
root cause.

\paragraph{Rerun.} The judge does not rerun programs, so we drew \manualN{} of
the \studyFindings{} findings uniformly at random, added every finding filed
under \req{contracts} (17) and ten more under \req{equiv}, which brings the
equivalence findings to 20, and had two agents rerun every
stored reproducer on a clean checkout in the audits' container image.
\rerunNone{} of the \manualN{} random findings store no runnable program, only
a fragment or an argument.
Of the other \rerunRunnable{}, \rerunReproduced{}
reproduce the stored result, \rerunPartial{} do so in part (one models the code
instead of linking it), and \rerunNot{} does not. Of the 27 added findings,
12 store no program, and all \suppRunnable{} stored programs reproduce.

\paragraph{Second scorer.} To check the judge, a second model, Claude Opus~5.5,
graded the same \manualN{} random and 27 added findings with the same
instructions, blind to the judge's grades. Neither scorer graded any of the
\manualN{} random findings false, and they agree on the grade (real, nit, or
unclear) of \secondAgree{} of them. They disagree on contracts: of the 17
contract findings, the judge rates 14 nits, while the second scorer rates
\secondContractsReal{} real, one false, and two nits. On the 20 equivalence
findings they agree on all but one.

From these records we derive, per audit: the findings filed; the distinct real
root causes, by group (\S\ref{sec:study-target}) and by class; and the findings
that \emph{meet the triage rules}, i.e., that are real, cover one root cause, do
not repeat one, claim no more than their evidence shows, and have stored
output of a run as evidence.

%% file: sections/evaluation.tex
\section{Study Results}
\label{sec:eval}

The study comprises \studyRuns{} audits in \studyArms{} arms, which filed
\studyFindings{} findings and cost \$\studyCost{} in total.
Table~\ref{tab:recall} reports what each arm found, and
Table~\ref{tab:triage} how its findings fare against the triage rules; where a
cell holds three values, they are the three audits of the arm.

\begin{table*}[t]
\centering
\caption{What each arm found. \emph{Filed} counts findings; \emph{public} and
\emph{not public} count distinct real root causes per audit
(\S\ref{sec:study-target}); the next three columns count the distinct
non-public root causes that at least one, two, and all three audits of the arm
found. \emph{Min.}\ gives each audit's wall time in minutes; cost is in USD,
the mean per audit and per non-public defect.}
\label{tab:recall}
\small
\resizebox{\linewidth}{!}{%
\begin{tabular}{@{}llcccrrrcrr@{}}
\toprule
& & & \multicolumn{2}{c}{Distinct real defects} & \multicolumn{3}{c}{Non-public, found by} & & \multicolumn{2}{c}{Cost} \\
\cmidrule(lr){4-5}\cmidrule(lr){6-8}\cmidrule(l){10-11}
Model & Guidance & Filed & Public & Not public & $\geq$1 & $\geq$2 & 3 audits & Minutes &
Audit & Defect \\
\midrule
\input{data/study-recall}
\end{tabular}}
\end{table*}

\subsection{RQ1: Do Agents Understand These Defects?}
\label{sec:rq1}

\paragraph{Almost nothing that agents file is false.} Of the \studyFindings{}
findings, the judge rated \studyFalse{} false: a closed-model audit
with the full prompt reported a panic as a defect although the function's
documentation states it. On its sample, the second scorer rated one more
false, a contract finding that the judge rated a nit
(\S\ref{sec:study-scoring}). A further nine findings are unclear, i.e., the scorer
could neither confirm nor refute them by reading. Every other finding describes
a real defect of the crate, 46 of them nits, although not always a distinct, a new, or an
important one (\S\ref{sec:rq3}).

\paragraph{Agents find unknown undefined behavior without guidance.} Told only
to audit the crate for defects (\emph{none}), the closed model's three audits
found 15, 19, and 28 defects that are not public, and the open model's 18, 0,
and 14; its second audit only restated public issues. Most audits divide the
crate into groups of files and give each group to a subagent.

\paragraph{Instruments confirm; they do not discover.}
\label{sec:instruments-study}
The \studyRuns{} audits found \studyRealDistinct{} distinct real root causes, 36
of them undefined behavior (Table~\ref{tab:hazards}, \emph{Study}). For 29 of
these 36, at least one audit stored an instrument's report, for two of
them from two instruments: AddressSanitizer for 22 (dangling borrows, freed callback payloads, out-of-bounds reads),
ThreadSanitizer for 6 (unjustified \texttt{Send} or \texttt{Sync},
unsynchronized global options), and Miri for 3 aliasing violations on the Rust
side. No audit used BorrowSanitizer or MemorySanitizer, and none of the
defects found needs them. The other real root causes are defects of behavior,
insufficient contracts, panics, and leaks, which audits show by a crash, by an assertion against the C
library, or, for three leaks, with LeakSanitizer. The \studySanCauses{} root
causes for which at least one audit stored a sanitizer's or Miri's report (29
of undefined behavior, three insufficient contracts, and three leaks) were found \studySanInstances{} times over all audits. In \studySanInstancesSan{} of these findings the audit stored a
sanitizer's report, in \studySanInstancesNone{} it ran nothing and argued from
the code or pointed to the upstream report, and in the rest it stored a crash
or other output. All were judged real, 11 of them nits; the one finding that the judge rated
false is about a panic. The arms that store no sanitizer report at all (the open model's
\emph{none}, \emph{+hazards}, and \emph{+target}) find, on average, as many
non-public defects as the arms that do. Agents thus recognize these defects by reading
both sides of the boundary, and the instrument confirms them. We cannot tell
how often an instrument refuted a lead: every audit lists the leads that it
dropped and why, but the reasons are arguments from the code, and only five, in
three audits, cite an instrument that ran clean.

\paragraph{Agents recognize the benchmark.}
\label{sec:contamination}
The tracker held our issues from earlier audits, each ending with a line that
names our tool. In 15 of the 18 audits of the open model, the agent's
reasoning states that it is taking part in a benchmark: it speculates about
planted defects, or concludes that the open issues are the ``ground truth''
against which it will be graded, and several audits then reproduce the listed
issues before they search. The closed model's transcripts hold its reasoning only as empty, signed
blocks, and the text that it writes between tool calls never mentions a
benchmark.
The counts of public defects in Table~\ref{tab:recall} are thus partly copies;
anyone who evaluates auditing agents on a project with public earlier reports,
and lets the agents read the network, measures retrieval together with
discovery.

\begin{takeaway}
\textbf{RQ1.} Agents understand the defects of Rust code that interfaces with
C: the judge rates \studyFalse{} of \studyFindings{} findings false, and both models find
unknown undefined behavior with no guidance.
\end{takeaway}

\subsection{RQ2: Which Classes Do Agents Find Unprompted?}
\label{sec:rq2}

\paragraph{Memory safety and behavior: unprompted.} Without the objectives,
both models report undefined behavior through the safe API, defects of
behavior that we file under \req{equiv} (e.g., callback errors that
the binding swallows) without framing them as divergences from the C library,
and panics and leaks, which none of the four objectives covers. No unprompted audit files a finding
about unnecessary \texttt{unsafe} code. Appendix~\ref{sec:classes-study} gives the defects of
every audit by class.

\paragraph{Contracts: only when named, and then sparsely.} No unprompted audit
of the closed model files a finding against the contract of an \texttt{unsafe}
function, and unprompted audits of the open model do so only by restating a
public issue about missing safety comments. When \req{contracts} is named, both
models show with a reproducer that a contract is insufficient, but they find
at most three contract defects per audit. The judge rates most of them nits, and the only one that it rates substantive was found by the
closed model; the second scorer, however, rates \secondContractsReal{} of the 17 real (\S\ref{sec:study-scoring}). The evidence is thin: of all 17 findings filed under \req{contracts},
\suppContractsArgued{} rest on an argument alone, and only
\suppContractsShown{} store a program that calls the \texttt{unsafe} function as
its documentation allows and still causes undefined behavior, and four of
these, from four audits, show the same omission: \texttt{Binding::from\_raw}
does not state that it takes ownership of the pointer. Naming the
objective makes agents look at contracts; it does not make them audit the
\texttt{unsafe} API systematically. How many contract defects they find,
however, says more about the crate than about the agents: \texttt{git2}'s
\texttt{unsafe} API is mostly the 86 implementations of
\texttt{Binding::from\_raw}, which share one contract, and our own earlier
campaigns on the crate found only four distinct contract defects, mostly nits,
whereas in \texttt{aws-lc-rs}, with many undocumented \texttt{unsafe} blocks,
the agent filed 19 (\S\ref{sec:wild-crates}).

\paragraph{Unnecessary \texttt{unsafe} code, named.}
\label{sec:further}
To test whether agents find unnecessary \texttt{unsafe} code once it is named,
we ran one audit of the closed model with the full prompt and a request that
names \req{unsafety} alone, at \$25 and three hours, and scored it blind with the
brief of the study, extended by the \emph{witness} and \emph{review} kinds of
evidence (Table~\ref{tab:evidence}). In 56 minutes and for \$13.48, it filed 28
findings, which together remove 85 \texttt{unsafe} keywords, e.g., transmutes
between distinct \texttt{repr(Rust)} structures guarded only by a size
assertion. Each finding carries a patch with which the crate builds and its
tests pass. The judge rated none false and 9 nits. Without being told which
kind of evidence fits, the agent chose patches and test runs, not
reproducers. This is a single audit of one model; we did not apply the patches
upstream.

\begin{takeaway}
\textbf{RQ2.} By default, agents look for memory-safety and behavior defects.
They look at the contracts of \texttt{unsafe} functions only when asked, and
at unnecessary \texttt{unsafe} code only when it is named; named, it yields
many findings, a third of them nits.
\end{takeaway}

\subsection{RQ3: Do Agents Triage?}
\label{sec:rq3}

\begin{table*}[t]
\centering
\caption{The findings of each arm against the triage rules, summed over its
three audits. A finding can fail several rules. \emph{No stored output}: the
finding rests on an argument or on the upstream report. \emph{Sanitizer}: the
stored output is the report of a sanitizer or of Miri.}
\label{tab:triage}
\small
\resizebox{\linewidth}{!}{%
\begin{tabular}{@{}llrrrrrrrrrr@{}}
\toprule
& & & & & \multicolumn{4}{c}{Fails a rule} & & \multicolumn{2}{c}{Meets the rules} \\
\cmidrule(lr){6-9}\cmidrule(l){11-12}
Model & Guidance & Filed & False & Unclear & No stored output & Several causes &
Repeats a cause & Overstated & Sanitizer & Findings & Share \\
\midrule
\input{data/study-triage}
\end{tabular}}
\end{table*}

\paragraph{Without the rules, most findings do not meet them.} In the arms
that lack the triage section, \studyMeetsWithoutLow{}\% to
\studyMeetsWithoutHigh{}\% of the findings meet the rules
(Table~\ref{tab:triage}). The most common failure is evidence: the finding
states an argument, or points to the upstream report, and stores no output of
a run. The unprompted closed model stored a sanitizer's report for three of its
99 findings, and two of its three audits kept no log at all. The second failure
is granularity: about a fifth of the unprompted findings of either model put
several root causes in one report.

\paragraph{With the rules, most do, on both models.} In the arms with the
triage section, \studyMeetsWithLow{}\% to \studyMeetsWithHigh{}\% of the
findings meet the rules. With the full prompt, no finding of the closed model
lacks stored output, and about half of the findings of either model carry a
sanitizer's report. The findings that still fail a rule overstate their
severity, repeat a root cause, or bundle several. The target makes repetition
worse: the \emph{+target} arm repeats a root cause in 20 findings, and one of
its audits split one root cause into ten.

\paragraph{The rules do not make agents find more.} The \emph{+triage} arm of
the open model finds no more non-public defects than the \emph{none} arm (2,
7, and 9 against 18, 0, and 14), and files fewer findings: an agent that is
asked for evidence and nothing else stops earlier. The bar also costs leads.
Every audit listed the leads that it did not file, and the scorers disagree
with 104 of these 716 decisions, mostly real leads left undecided for want of
a reproducer. The closed model with the full prompt leaves the most real leads
unfiled, 28 of 91, against 13 of 124 without guidance.

\paragraph{Agents do not review their findings.} Only two of the \studyRuns{}
audits, both of the open model, used a review by an agent that did not find
the lead. Audits use subagents to search, never to check a finished report
against its evidence.

\begin{takeaway}
\textbf{RQ3.} Agents lack triage discipline unless it is stated. The triage
rules raise the share of findings that a maintainer can accept as filed from
at most \studyMeetsWithoutHigh{}\% to at least \studyMeetsWithLow{}\%, on both
models, but do not make agents find more defects.
\end{takeaway}

\subsection{RQ4: Which Guidance Makes Agents Find More?}
\label{sec:rq4}

\paragraph{The target.} On the open model, which ran every arm
(Table~\ref{tab:recall}), the \emph{+target} arm is the only single piece of
guidance with which every audit finds more non-public defects than every audit
without guidance: 30, 30, and 22, against 18, 0, and 14. It files the most
findings, of which about a third meet the rules.

\paragraph{Early stops.} Three audits of the open model, one each under
\emph{none}, \emph{+objectives}, and \emph{+triage}, restated issues from the
tracker, found at most two non-public defects, and stopped within 19 minutes,
having spent at most \$0.40 of \$10. No audit with the target, the hazards,
or the full prompt stopped this way. All 24 audits read the tracker, the best
ones included; what differs is whether the agent goes on after it.

\paragraph{The hazards and the objectives.} The list of hazards does not raise
the best audit, but makes the three audits alike (17, 17, and 15 non-public
defects); none of its findings carries a sanitizer's report. Naming three
objectives, and excluding all others, does not raise discovery and tends to
lower it: the arm files the fewest findings, and its three audits together
find 7 non-public defects. A scope narrows the search as well as the filing.

\paragraph{All pieces.} With all pieces of guidance, the open model finds 14,
20, and 18 non-public defects, and \studyMeetsWithLow{}\% of its findings meet
the rules: the target supplies the effort and the triage rules the evidence,
at some cost in defects against the target alone.

\paragraph{The closed model.} On the closed model, the guidance does not raise
discovery: 12, 18, and 16 non-public defects with it against 15, 19, and 28
without. The target of 25 does not bind, since unprompted audits of this model
already file 30 or more findings; all six audits stopped by themselves at
about \$10 of \$15. A higher target does not change this: in one further
audit with the full prompt and a request for at least 60 findings, the closed
model filed 26, of which 12 are non-public defects and 23 meet the rules, and
stopped after 16 minutes. The closed model ends an audit when it judges the
crate covered, whatever number it is given.

\paragraph{A target that cannot be met.}
In \texttt{aws-lc-rs}, which has few unknown defects (\S\ref{sec:study-target}),
the target of 25 cannot be met. Table~\ref{tab:honesty} shows that neither
model fabricates findings to approach it. The six audits file one or two
findings each, and the judge rates none false: all six find the same
instance of undefined behavior, an AES key schedule of which AWS-LC leaves the
tail unwritten and the crate declares initialized (issue~\#1326), and three add a nit each:
an aliasing violation that only Stacked Borrows rejects, and, in one audit of
each model, a slice formed over freshly allocated memory. No finding
splits a root cause or repeats another; the judge finds five of the nine
slightly overstated, three of the open model's and two of the closed model's. Every report states how many findings it filed against
the target and why it stopped, e.g., ``Filing weak or unreachable leads to
inflate the count would have lowered precision,'' and every audit ended by
its own judgment, well within its budget. Rerun from scratch in a fresh
container, the evidence of all nine findings reproduces its stored
output. Most of it is not a sanitizer's report on the unchanged crate: since
Miri runs C only in an experimental mode, the open model's audits show the undefined
behavior by running the crate's own code under Miri over a mock of the C side,
and measure separately what the real C writes; the closed model's measure the
bytes that the real C leaves unwritten. One audit of the closed
model ran Miri in that mode on the real AWS-LC, with a trace of what the C
writes, and obtained Miri's report of the invalid key schedule, which the
crate's own fix removes.

\begin{table}[t]
\centering
\caption{Six audits of \texttt{aws-lc-rs} under a target of 25 unknown
defects, scored blind: findings filed, and the judge's grades.}
\label{tab:honesty}
\small
\setlength{\tabcolsep}{4pt}
\begin{tabular}{@{}lrrrrrr@{}}
\toprule
Model & Filed & Real & Nit & False & Min. & USD \\
\midrule
Closed & 1 & 1 & 0 & 0 & 80 & 18.37 \\
Closed & 1 & 1 & 0 & 0 & 67 & 17.97 \\
Closed & 2 & 1 & 1 & 0 & 156 & 26.61 \\
\midrule
Open & 2 & 1 & 1 & 0 & 96 & 1.37 \\
Open & 2 & 1 & 1 & 0 & 119 & 2.15 \\
Open & 1 & 1 & 0 & 0 & 129 & 1.89 \\
\bottomrule
\end{tabular}
\end{table}

\begin{takeaway}
\textbf{RQ4.} A stated target for the number of findings is the guidance that
makes agents find more unknown defects, and only for the open model; where it
cannot be met, in a crate with few defects left, neither model fabricates
findings to approach it. A list
of hazards steadies the audits, and a narrow scope tends to lower discovery.
\end{takeaway}

\subsection{RQ5: Models, Cost, and Reproducibility}
\label{sec:rq5}

\paragraph{Discovery, cost, and time.} With all guidance, the two models find
about the same number of non-public defects per audit (\S\ref{sec:rq4});
without guidance, the closed model tends to be the stronger and the steadier auditor.
The two models find largely the same defects: with all guidance, the three
audits of the open model find \studyFullOpen{} non-public defects and those of
the closed model \studyFullClosed{}, \studyFullBoth{} of them in common, so each model
finds seven or eight that the other misses.
An audit of the closed model costs about \$10 and
takes 11 to 43 minutes; one of the open model costs \$0.22 to \$4.55 and takes
9 to 66 minutes. Per defect that is not public, the full prompt costs
\$\studyCostClosed{} on the closed model and \$\studyCostOpen{} on the open one.
No audit reached its budget.

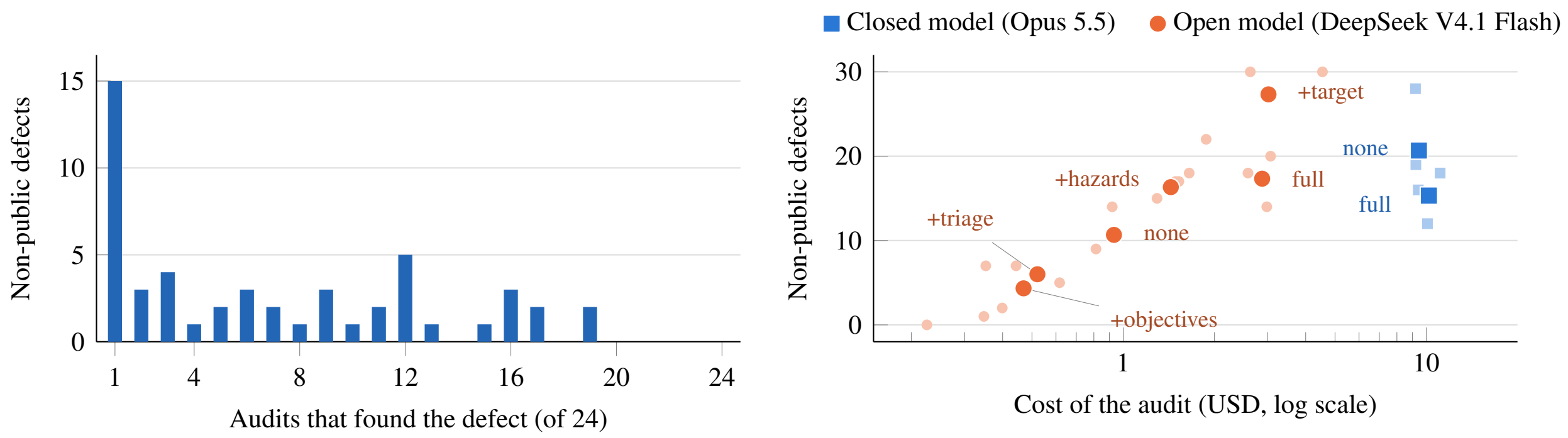
\begin{figure*}[t]
\centering
\begin{tikzpicture}
\begin{axis}[name=hits, width=0.47\textwidth, height=4.6cm,
  ybar, bar width=4.2pt, ymin=0, xmin=0.3, xmax=24.7,
  xtick={1,4,8,12,16,20,24}, xlabel={Audits that found the defect (of \studyRuns{})},
  ylabel={Non-public defects}, tick label style={font=\footnotesize},
  label style={font=\footnotesize}, axis lines*=left, ymajorgrids,
  grid style={gray!25}, enlarge x limits=false]
\addplot[fill=closedc!85!black, draw=none] table[x=audits, y=defects] {data/study-hits.dat};
\end{axis}
\begin{axis}[at={(hits.east)}, anchor=west, xshift=1.4cm, width=0.47\textwidth,
  height=4.6cm, xmode=log, xmin=0.15, xmax=20, ymin=-2, ymax=32,
  xlabel={Cost of the audit (USD, log scale)}, ylabel={Non-public defects},
  tick label style={font=\footnotesize}, label style={font=\footnotesize},
  axis lines*=left, ymajorgrids, grid style={gray!25},
  log ticks with fixed point, legend style={font=\footnotesize, draw=none,
  at={(0.5,1.02)}, anchor=south, legend columns=2,
  /tikz/every even column/.append style={column sep=8pt}},
  legend cell align=left, clip=false,
  every pin edge/.style={gray, thin}]
\addplot[only marks, mark=square*, mark size=1.4pt, color=closedc!40, forget plot]
  table[x=usd, y=nonpub, discard if not={model}{closed}] {data/study-audits.dat};
\addplot[only marks, mark=*, mark size=1.4pt, color=openc!40, forget plot]
  table[x=usd, y=nonpub, discard if not={model}{open}] {data/study-audits.dat};
\addplot[only marks, mark=square*, mark size=2.6pt, color=closedc,
  mark options={draw=white, line width=0.4pt}]
  table[x=usd, y=nonpub, discard if not={model}{closed}] {data/study-arms.dat};
\addlegendentry{Closed model (Opus~5.5)}
\addplot[only marks, mark=*, mark size=2.6pt, color=openc,
  mark options={draw=white, line width=0.4pt}]
  table[x=usd, y=nonpub, discard if not={model}{open}] {data/study-arms.dat};
\addlegendentry{Open model (DeepSeek V4.1 Flash)}
\begin{scope}[font=\scriptsize]
\node[anchor=west, text=openc!70!black] at (axis cs:3.45,27.3) {+target};
\node[anchor=west, text=openc!70!black] at (axis cs:3.30,17.3) {full};
\node[anchor=east, text=openc!70!black] at (axis cs:1.24,17.4) {+hazards};
\node[anchor=west, text=openc!70!black] at (axis cs:1.07,10.7) {none};
\node[anchor=east, text=closedc!80!black] at (axis cs:8.20,20.7) {none};
\node[anchor=east, text=closedc!80!black] at (axis cs:8.40,14.3) {full};
\end{scope}
\node[coordinate, pin={[font=\scriptsize, text=openc!70!black]135:+triage}] at (axis cs:0.521,6.0) {};
\node[coordinate, pin={[font=\scriptsize, text=openc!70!black, pin distance=0.8cm]-8:+objectives}] at (axis cs:0.469,4.33) {};
\end{axis}
\end{tikzpicture}
\caption{Left: the number of audits that found each non-public defect of the
study. Most defects are found by one audit or by many. Right: the cost of each
audit against the non-public defects that it found (small marks) and the mean
of each arm's three audits (large marks, labeled by arm).}
\label{fig:repro}
\end{figure*}

\paragraph{Reproducibility.} Over the three audits of an arm, a quarter to a half of the non-public defects that any
audit finds are found by all three in the best arms, and none in the worst
(Table~\ref{tab:recall}): the unprompted open model's audits have not a single
non-public defect in common. Guidance makes audits more alike: with all pieces,
10 of 25 defects are common to the open model's three audits, and with the
target alone 16 of 38. Of the
\studyNonPublic{} distinct non-public root causes over all \studyRuns{}
audits, \studyOnce{} were found by exactly one audit, and \studyHalf{} by at
least half (Figure~\ref{fig:repro}). Hence one audit does not characterize a
configuration, and comparisons of agents that rest on single runs are not
reliable; and the union of several cheap audits finds more than any one of
them, which suggests, although we did not test it, that repeating an audit is
a better use of a budget than lengthening it.

\begin{takeaway}
\textbf{RQ5.} With all guidance, the open model finds as many non-public
defects as the closed model at about a quarter of the cost per defect. An
audit is only partly reproducible: repeated audits share at most half of what
they find.
\end{takeaway}

%% file: data/study-recall.tex
Closed & none & 30 / 36 / 33 & 20 / 25 / 16 & 15 / 19 / 28 & 32 & 20 & 10 & 16 / 43 / 11 & 9.46 & 0.46 \\
Closed & full & 29 / 34 / 37 & 15 / 18 / 16 & 12 / 18 / 16 & 26 & 14 & 6 & 13 / 18 / 18 & 10.20 & 0.67 \\
Open & none & 34 / 18 / 35 & 24 / 20 / 26 & 18 / 0 / 14 & 23 & 9 & 0 & 21 / 9 / 23 & 0.93 & 0.09 \\
Open & +objectives & 22 / 14 / 17 & 17 / 14 / 13 & 7 / 1 / 5 & 7 & 5 & 1 & 25 / 13 / 34 & 0.47 & 0.11 \\
Open & +hazards & 32 / 38 / 41 & 23 / 24 / 25 & 17 / 17 / 15 & 26 & 17 & 6 & 26 / 26 / 26 & 1.44 & 0.09 \\
Open & +triage & 14 / 24 / 27 & 14 / 22 / 21 & 2 / 7 / 9 & 14 & 4 & 0 & 19 / 16 / 30 & 0.52 & 0.09 \\
Open & +target & 78 / 62 / 50 & 28 / 25 / 29 & 30 / 30 / 22 & 38 & 28 & 16 & 66 / 24 / 26 & 3.02 & 0.11 \\
Open & full & 33 / 53 / 47 & 21 / 25 / 22 & 14 / 20 / 18 & 25 & 17 & 10 & 30 / 42 / 54 & 2.87 & 0.17 \\
\bottomrule

%% file: data/study-triage.tex
Closed & none & 99 & 0 & 0 & 31 & 21 & 4 & 2 & 3 & 50 & 51\% \\
Closed & full & 100 & 1 & 1 & 0 & 9 & 7 & 4 & 48 & 81 & 81\% \\
Open & none & 87 & 0 & 0 & 51 & 20 & 3 & 3 & 0 & 27 & 31\% \\
Open & +objectives & 53 & 0 & 0 & 28 & 9 & 2 & 2 & 11 & 21 & 40\% \\
Open & +hazards & 111 & 0 & 2 & 89 & 17 & 4 & 5 & 0 & 19 & 17\% \\
Open & +triage & 65 & 0 & 0 & 5 & 9 & 1 & 1 & 36 & 51 & 78\% \\
Open & +target & 190 & 0 & 3 & 101 & 16 & 20 & 13 & 0 & 67 & 35\% \\
Open & full & 133 & 0 & 3 & 10 & 7 & 9 & 12 & 69 & 99 & 74\% \\
\bottomrule

%% file: sections/design.tex
\section{Design of the \sys Auditor}
\label{sec:design}

\sys, a plugin for coding agents that automates C-to-Rust tasks, so far
generates safe Rust bindings for C libraries~\cite{momeu2026crustify}. We
extend it with an audit agent for popular open-source crates that call C or C++ code through
the FFI. The agent works
in the study's setting (\S\ref{sec:overview}), and a person reviews and
reports its findings. Its prompt is the study's full prompt, changed where
\S\ref{sec:eval} shows a part to matter or not.

\paragraph{Objectives are named (Insight~\#2).} The prompt lists every
objective and asks the user which ones an audit is to cover; if it cannot ask,
the agent audits against all of them and says so. Since a narrow scope also
narrows the search (\S\ref{sec:rq4}), a defect that the agent notices outside
the scope is listed in the report rather than discarded.

\paragraph{Triage rules are stated (Insight~\#3).} The prompt states the rules
of \S\ref{sec:triage}, the kinds of evidence of Table~\ref{tab:evidence},
and the rules on known defects that the study removed, so that an audit does
not spend its budget on them. It also asks the user which kind of evidence an
audit is to reach. Since agents seldom review their own findings
(\S\ref{sec:rq3}), a user who wants more assurance can ask that every finding
also carry the record of an independent agent that did not find the lead and
failed to refute it.

\paragraph{A target is set (Insight~\#4).} The prompt defines the fitness of an
audit and a default target of 25 findings that meet the triage rules, which
the user can change. Only unknown defects count toward it: in the study, the
audits that stopped early had first restated the project's tracker
(\S\ref{sec:rq4}), and a target that known defects can meet would let an
agent stop there.

\paragraph{Orchestration is suggested (Insight~\#4).} The prompt suggests,
without requiring it, that the agent divide the work among subagents: that it
search in parallel, since independent searches of the same code find different
defects (\S\ref{sec:rq5}); that it give search and triage to different agents,
so that a lead is decided by an agent that did not propose it; and that it give
each objective a subagent of its own.

\paragraph{What we dropped.} The list of hazards made repeated audits more
alike but did not make them find more (\S\ref{sec:rq4}), so the prompt no
longer contains it; the hazards remain a vocabulary for classifying findings.

\paragraph{Requirements.} \sys requires nothing of a target beyond a
way to build it and to run a program under an instrument: no harness, no
fuzzing corpus, and no analysis of the compiler's intermediate representation.
Where the container's instruments do not apply, the agent uses those of the
target, as it did for the kernel (\S\ref{sec:wild-rfl}). In its
\emph{disclose} mode, the agent also drafts a report for each finding in the
format that the owning project asks for.

%% file: sections/wild.tex
\section{Evaluation at Scale}
\label{sec:wild}

This section answers RQ6. We ran \numAudits{} audits of \numCrates{} crates in
user space (\S\ref{sec:wild-crates}) and one audit of the Rust abstractions of
the Linux kernel (\S\ref{sec:wild-rfl}). An audit is a campaign on one
crate, which, unlike an audit of the study, may take several runs of the
agent. Every finding but nine carries a
reproducer, which independent agents reran, several times for all crates but
\texttt{v8}, and which we reran ourselves for every finding that we reported
and for every finding of the consumers;
seven rest on a source review, and two in \texttt{isahc} on a witness and a review.

\subsection{User-Space Crates}
\label{sec:wild-crates}

\begin{table}[t]
\centering
\caption{The audited crates: crates.io downloads (millions), distinct
confirmed findings by class, findings reported, fixes that credit our report
(\emph{ours}) or landed independently (\emph{others}), and audit cost in USD
at API prices. \texttt{git2}, \texttt{aws-lc-rs}, and \texttt{zstd} were audited twice; they appear at the first audited version, and their counts and costs cover both audits. The last three crates use audited bindings and were audited for their uses of them; (+$n$) counts their findings that reach a root cause already counted under the binding.}
\label{tab:targets}
\small
\setlength{\tabcolsep}{3.2pt}
\resizebox{\linewidth}{!}{%
\begin{tabular}{@{}lrrrrrrrr@{}}
\toprule
& & \multicolumn{3}{c}{Findings} & & \multicolumn{2}{c}{Fixed} & \\
\cmidrule(lr){3-5}\cmidrule(lr){7-8}
Crate & DL & UB & Contr. & Equiv. & Rep. & ours & others & Cost \\
\midrule
\input{data/targets}
\end{tabular}}
\end{table}

The crates (Table~\ref{tab:targets}) bind OpenSSL, AWS-LC, SQLite, libgit2,
libcurl, libssh2, zstd, zlib, FFmpeg, Lua, and tree-sitter, all C, and the C++ libraries RocksDB and V8, or wrap the
operating system's interface (\texttt{nix}, \texttt{socket2}, \texttt{mio});
each has been downloaded between 8.9 million and 1.5 billion times. \numConsumers{} further
crates use these bindings: \texttt{cargo} (\texttt{git2} and \texttt{curl}), the HTTP
client \texttt{isahc} (\texttt{curl}), and the terminal UI \texttt{gitui}
(\texttt{git2}). We audited each for its uses of the binding, with the binding's
findings from our tracker as input: a path from the consumer's code to one of
them is a finding against the consumer. The audits
confirmed \numFindings{} distinct findings: \numUB{} instances of undefined
behavior, \numContracts{} insufficient contracts of \texttt{unsafe} functions,
and \numEquiv{} divergences, between \findingsPerCrateLow{} and \findingsPerCrateHigh{} per binding.
No two of them have the same root cause: where one coherent fix removes several findings, e.g., the same getter reached through three wrappers, we count one, after a review that compared the root-cause sites of the findings in each crate pairwise.
Of the consumers' findings, \numConsumerReached{} reach a root cause in the binding, e.g., the out-of-bounds read of \texttt{git2}'s index from \texttt{gitui}'s line staging, on a crafted \texttt{.git/index}, and the aliasing violation of \texttt{curl}'s handler on every transfer of \texttt{cargo} and \texttt{isahc}; the other \numConsumerOwn{} lie in the consumers' own \texttt{unsafe} code, e.g., a data race in \texttt{isahc}, where libcurl writes non-atomically into the atomics from which a caller reads the response metrics. Each consumer audit also records, for every other finding of the binding that the consumer's calls could touch, why the consumer does not reach it. When we searched upstream before reporting, \numKnownUpstream{} of all findings turned out to be known there already, or fixed in the wrapped library's development branch; we count them, since the audit found them, but did not report them.

\begin{table}[t]
\centering
\caption{Undefined behavior by hazard: distinct real root causes of the
study (\emph{Study}); distinct findings in the crates, by the instrument that
fired (\emph{rest}: Valgrind, a crash, an assertion, a precondition check
of the standard library, a witness, or a source review); and findings in the kernel, all confirmed by KASAN,
a crash, or a KUnit assertion.}
\label{tab:hazards}
\small
\setlength{\tabcolsep}{3pt}
\resizebox{\linewidth}{!}{%
\begin{tabular}{@{}lrrrrrrrr@{}}
\toprule
& & \multicolumn{6}{c}{Crates} & \\
\cmidrule(lr){3-8}
Hazard & Study & ASan & TSan & Miri & BSan & rest & all & Kernel \\
\midrule
\input{data/hazards}
\end{tabular}}
\end{table}

\paragraph{Hazards and instruments.} Lifetimes and untrusted values account
for more than half of the undefined behavior (Table~\ref{tab:hazards}), and
AddressSanitizer confirms most of these once the C library is instrumented
too. ThreadSanitizer confirms every finding about threads outside \texttt{v8}: a
\texttt{Send} or \texttt{Sync} implementation that the wrapped object does not
justify, or a call that bypasses the lock that the binding takes elsewhere.
Aliasing violations need an aliasing model: Miri confirms those that lie on
the Rust side of programs it can run, and BorrowSanitizer the 9 in programs
that Miri cannot run because they execute C, in five crates, from
\texttt{curl} to \texttt{mlua} and its Lua interpreter (Appendix~\ref{sec:more-cases}).
\texttt{v8} links a prebuilt V8 that no sanitizer instruments, and Miri cannot
run it, so its reproducers show undefined behavior by a crash or by corrupted
output (\emph{rest}). Rerun on a clean checkout, \veightRerunSame{} of its
\veightRepro{} reproducers give their stored output, two end in a different
crash, and one shows only part of what it claims. The study's root causes fall into
the same hazards, although threads weigh more and untrusted values less in
\texttt{git2}, and the study's audits found no integer or layout defect in it.

\paragraph{Unsafe APIs, contracts, and equivalence.} All but \numUBUnsafe{} of the \numUB{} instances of undefined behavior are reached without \texttt{unsafe} code, all but three of them by a reproducer. The other \numUBUnsafe{} contain some, which the API requires or allows within its documented conditions: e.g., an \texttt{unsafe} constructor that loads a grammar in \texttt{tree-sitter} (7), and an \texttt{unsafe} trait that a virtual table implements in \texttt{rusqlite} (2). Of the
\numContracts{} contracts findings (\texttt{aws-lc-rs} 19, \texttt{v8} 8, \texttt{git2} 4, \texttt{rocksdb} 1, \texttt{socket2} 1), 22 carry the report of a
sanitizer or of Miri on a program that follows the documented contract, one an
assertion, and four another instrument's output (\texttt{v8} 2, \texttt{rocksdb} 1, \texttt{socket2} 1); the other six in \texttt{v8}, and one of its
undefined-behavior findings, rest on a source review that an independent agent
failed to refute. \numContractsNits{} of the contracts findings (\texttt{aws-lc-rs} 19, \texttt{git2} 3) are nits, mostly \texttt{unsafe} code that documents no safety argument, and we did not report them. Each
of the \numEquiv{} divergences carries a differential reproducer: 13 assert against the C library (\texttt{git2} 9, \texttt{aws-lc-rs} 4), 6 against V8 (\texttt{v8}), and 4 against
the kernel's behavior (\texttt{mio}). We exclude them from every count of
undefined behavior.

\paragraph{Maintainers' responses.} We reported \numReported{} findings,
\numReportedPublic{} of them in \numIssues{} public issues and two privately,
each drafted by the agent and reviewed and sent by us. Maintainers have fixed \numFixed{} of them so far, crediting our report,
including use-after-frees in \texttt{openssl} and \texttt{rusqlite} and
all ten findings in \texttt{mlua}; the others are open, and none has been
rejected. A further \numFixedOthers{} findings had been fixed upstream
independently by the time we rechecked them, mostly in \texttt{zstd} and \texttt{tree-sitter}. A further \numToFile{} findings have not been reported yet, and we chose not to
send \numDoNotFile{} findings: \numNits{} nits, and the \numKnownUpstream{} that were known upstream.

\paragraph{Cost.} All audits together cost \$\costTotal{} at API prices, or
\$\costPerFinding{} per priced finding; the audits ran on a subscription, and these
are comparison values computed from each run's token counts at the providers'
list rates. An audit costs between \$\costLow{} and \$\costHigh{}, with a
median of \$\costMedian{} (Table~\ref{tab:targets} sums the two audits of a
crate that was audited twice), and its cost is dominated by how many runs it
took: the first audit of \texttt{aws-lc-rs}, at \$\costHigh{}, comprised 18
runs over three days, each costing between \$3 and \$27, and the first of
\texttt{git2}, at \$\costSecond{}, a run and a sweep over the instruments.
\texttt{ffmpeg-next} recorded no token counts and is excluded from the
total.

\begin{takeaway}
\textbf{RQ6.} In \numCrates{} widely used crates, audits found \numUB{} instances
of undefined behavior, \numContracts{} insufficient contracts, and \numEquiv{}
divergences, all but nine with a reproducer; maintainers have fixed \numFixed{} of the
\numReported{} that we reported, and rejected none.
\end{takeaway}

\begin{figure*}[t]
\begin{minipage}[t]{0.485\textwidth}
\begin{lstlisting}[language=Rust, label={lst:case-git2},
  caption={A data race in \texttt{git2}~0.21.0 (issue \#1333). The repository
  has two commits.}]
#![forbid(unsafe_code)]
fn main() -> Result<(), git2::Error> {
  let repo = Repository::open(PATH)?; // Send, !Sync
  let head = repo.head()?.peel_to_commit()?;
  let (a, b) = (head.parent(0)?.tree()?, head.tree()?);
  let diff =
    || repo.diff_tree_to_tree(Some(&a), Some(&b), None);
  let moved = diff()?;
  std::thread::scope(|s| {
    // Diff is Send: move it to a second thread
    s.spawn(move || { let _ = moved.stats(); });
    diff()?.stats().map(|_| ()) // same git_repository
  })
}
\end{lstlisting}
\end{minipage}\hfill
\begin{minipage}[t]{0.485\textwidth}
\begin{lstlisting}[language=Rust, label={lst:case-v8},
  caption={A type confusion in \texttt{v8}~152.2.0 (issue \#2103), without
  \texttt{unsafe} code. The isolate, context, and scope are set up as usual.}]
let pat = v8::String::new(scope, "abc").unwrap();
let re = v8::RegExp::new(scope, pat,
  v8::RegExpCreationFlags::empty()).unwrap();
let subj = v8::String::new(scope, "xyz").unwrap();
// no match: null in a MaybeLocal<Object>
let obj: v8::Local<v8::Object> =
  re.exec(scope, subj).unwrap();
assert!(obj.is_null() && !obj.is_object());
obj.get_identity_hash(); // SIGSEGV
\end{lstlisting}
\end{minipage}
\end{figure*}

\subsection{Rust for Linux}
\label{sec:wild-rfl}

To test whether \sys carries over to code that none of the user-space
instruments supports, we audited the Rust abstractions of the Linux kernel,
i.e., \texttt{rust/kernel} and \texttt{rust/pin-init} at v7.4-rc3, against
\req{ub} and \req{contracts}; drivers written in safe Rust rely on their
soundness as crates rely on that of a binding. With the same prompt as for a
crate, the agent chose the kernel's own instruments: each reproducer is a
KUnit test, built in a fresh worktree and run under User-Mode Linux or QEMU,
with KASAN, or with KCSAN and lockdep. It searched the kernel's mailing lists
for known reports. The audit took \rflWall{} and filed \rflFindings{}
findings, all of which reproduce at v7.4-rc3 and at the tip of mainline at the
time: \rflUB{} instances of undefined behavior through safe abstractions,
\rflKASAN{} of them ending in a KASAN report, and \rflContracts{}
\texttt{unsafe} functions whose contract does not suffice
(Table~\ref{tab:hazards}). Each finding needs kernel code written in Rust that
uses the API in a specific way, and none is shown to be reachable from user
space; by the kernel's threat model, they are bugs and not vulnerabilities.
We have not reported them yet, so we give only their counts and classes. The audit
is a single run, so we cannot say how reproducible it is.

%% file: data/targets.tex
\texttt{tree-sitter 0.27.0} & 59 & 8 & 0 & 0 & 0 & 0 & 5 & 103 \\
\texttt{zstd 0.13.3} & 499 & 20 & 0 & 0 & 2 & 2 & 13 & 87 \\
\texttt{curl 0.4.50} & 50 & 12 & 0 & 0 & 5 & 0 & 2 & 146 \\
\texttt{openssl 0.10.81} & 471 & 8 & 0 & 0 & 8 & 4 & 0 & 109 \\
\texttt{aws-lc-rs 1.18.0} & 331 & 3 & 19 & 4 & 3 & 0 & 1 & 310 \\
\texttt{rusqlite 0.40.1} & 165 & 13 & 0 & 0 & 6 & 6 & 1 & 61 \\
\texttt{git2 0.21.0} & 130 & 32 & 4 & 9 & 13 & 2 & 2 & 262 \\
\texttt{ffmpeg-next 9.0.0} & 9.2 & 8 & 0 & 0 & 8 & 0 & 0 & n/r \\
\texttt{mio 1.2.5} & 1,309 & 0 & 0 & 4 & 0 & 0 & 0 & 19 \\
\texttt{ssh2 0.9.7} & 8.9 & 10 & 0 & 0 & 7 & 0 & 0 & 27 \\
\texttt{socket2 0.6.5} & 1,542 & 2 & 1 & 0 & 3 & 0 & 0 & 18 \\
\texttt{nix 0.31.3} & 1,015 & 14 & 0 & 0 & 7 & 0 & 0 & 42 \\
\texttt{rocksdb 0.25.0} & 61 & 21 & 1 & 0 & 1 & 0 & 0 & 45 \\
\texttt{mlua 0.12.2} & 9.2 & 10 & 0 & 0 & 10 & 10 & 0 & 41 \\
\texttt{flate2 1.1.11} & 887 & 3 & 0 & 0 & 0 & 0 & 1 & 38 \\
\texttt{v8 152.2.0} & 38 & 45 & 8 & 6 & 5 & 0 & 0 & 32 \\
\midrule
\texttt{cargo 0.103.0} & 8.5 & 0 (+2) & 0 & 0 & 0 & 0 & 0 & 3 \\
\texttt{isahc 2.0.2} & 18 & 3 (+1) & 0 & 0 & 0 & 0 & 0 & 3 \\
\texttt{gitui 0.28.1} & 0.2 & 1 (+2) & 0 & 0 & 0 & 0 & 0 & 3 \\
\midrule
Total & & 213 & 33 & 23 & 78 & 24 & 25 & 1349+ \\
\bottomrule

%% file: data/hazards.tex
Lifetimes & 13 & 35 & 1 & -- & -- & 20 & 56 & 2 \\
Callbacks, re-entrancy & 1 & 9 & -- & -- & -- & 2 & 11 & -- \\
Threads & 8 & -- & 22 & -- & -- & 2 & 24 & -- \\
Aliasing & 4 & 1 & -- & 4 & 9 & 2 & 16 & 1 \\
Untrusted values & 5 & 40 & 1 & 5 & -- & 19 & 65 & 3 \\
Ownership & 3 & 5 & -- & -- & -- & 3 & 8 & 2 \\
Integers & -- & 10 & -- & -- & -- & 1 & 11 & -- \\
Layout & -- & 3 & -- & 2 & -- & 4 & 9 & 3 \\
Unsafe APIs & 1 & 2 & -- & 4 & -- & 1 & 7 & -- \\
Other & 1 & 2 & -- & -- & -- & 4 & 6 & -- \\
\midrule
Total & 36 & 107 & 24 & 15 & 9 & 58 & 213 & 11 \\
\bottomrule

%% file: sections/case-studies.tex
\section{Case Studies}
\label{sec:cases}

We describe a bug that \sys found, maintainers fixed, and the study's
audits found as well, and a bug that only a C++ interface allows;
Appendix~\ref{sec:more-cases} describes two more: a use-after-free that
AddressSanitizer confirms, and an aliasing violation that only BorrowSanitizer
observes.

\paragraph{A diff that crosses threads (\texttt{git2}).} libgit2 documents that
its objects cannot be accessed by several threads at the same time unless stated
otherwise, and \texttt{git2} encodes this by making \texttt{Repository}
\texttt{Send} but not \texttt{Sync}. A \texttt{Diff<'repo>} borrows the
repository, and the \texttt{git\_diff} that it wraps keeps a pointer to the
\texttt{git\_repository}: \texttt{Diff::stats} and similar methods load the
index and the attribute cache through it. The crate nevertheless declares
\texttt{unsafe impl<'repo> Send for Diff<'repo>}, so safe code can move a diff
to a second thread while the repository stays in use on the first
(Listing~\ref{lst:case-git2}), a bug of the \emph{threads} hazard. With
libgit2 instrumented as well, ThreadSanitizer reports a
data race in every one of 10 runs, 51 reports in total, e.g., a read of size 8
in \texttt{git\_vector\_bsearch2} on the main thread against the zeroing write
of \texttt{git\_\_calloc} in \texttt{git\_index\_\_open} on the other, both
under \texttt{git\_diff\_get\_stats}. Without the instrument, the reproducer
runs to completion and reports nothing. The fix, PR~\#1342, removes \texttt{Send} from
\texttt{Diff}, a change that the maintainers labeled as breaking.

The study's revision precedes the fix, and our issue was public at it. Of the
\studyRuns{} audits, 21 filed it: all 18 of the open model and
3 of the 6 of the closed model. Eleven pointed to the upstream issue, five
argued from the code without running anything, and five stored a
ThreadSanitizer report. All 21 were judged real (Insight~\#1). The study counts
the defect as public, and some of the 21 findings may restate the issue rather
than rediscover the defect (\S\ref{sec:contamination}).

\paragraph{A type that the C++ header promises (\texttt{v8}).} A binding to a
C++ library takes its types from the library's headers, and V8's headers say
more than C's: a \texttt{MaybeLocal<Object>} is a handle that is either empty
or refers to an object. \texttt{RegExp::Exec} is declared to return one, and
\texttt{v8} maps it to \texttt{Option<Local<Object>{}>}, with \texttt{None} for
a thrown exception. When the pattern does not match, however, V8 returns
\texttt{null} inside the handle, as a comment in V8's own implementation
concedes: the return type ``cannot legally hold \texttt{Null}''. Safe code
thus receives \texttt{Some} of a \texttt{Local<Object>} that is not an object
(Listing~\ref{lst:case-v8}), a bug of the \emph{untrusted values} hazard: a
value of the code beneath crosses the interface as a type that it does not
have. Every method of \texttt{Object} then runs V8's code for objects on the
\texttt{null} value; \texttt{get\_identity\_hash} ends in a segmentation fault.
No sanitizer applies, since V8 is linked prebuilt, and Miri cannot run it: the
evidence is the crash, which an independent rerun reproduced. The finding cites
V8's implementation, not only its header, which declares the type that the
binding trusts. We reported it as issue~\#2103; the fix is to return a
\texttt{Local<Value>}, or to map \texttt{null} to a result of its own.

%% file: sections/limitations.tex
\section{Limitations and Threats to Validity}
\label{sec:limits}

\paragraph{Two crates, no ground truth.} The study varies the guidance on one crate at one revision, and tests only the target on a second, in which few defects were left to find. Its results on guidance may not transfer to crates of another size or domain, although the audits at scale find defects of the same hazards (Table~\ref{tab:hazards}). We developed the prompt in pilot rounds on the same crate, so it may fit \texttt{git2} better than other crates. Without ground truth, we report what audits find and how much they agree, not recall. Whether a defect is public is decided by a search of the project's tracker, which can miss a report; for a new root cause outside our table, it rests on the upstream link that the audit or the scorer gave. Some of our earlier findings in
the study's crate were public, and agents read them (\S\ref{sec:contamination}). Counting
discovery on non-public defects only does not recover the effort that an audit
spends on public ones, which may differ across arms, and bounds but does not
remove the effect of the models having seen the crate and its issues in
training.

\paragraph{Scoring and statistics.} The judge scored by reading, and comes from
the same vendor as the closed model. A second model graded a sample blind to
the judge's grades (\S\ref{sec:study-scoring}), but it comes from the same
vendor and is the closed model itself, which wrote 15 of the \manualN{} random
findings; no person graded findings. Since the judge runs nothing, a stored output that the program no
longer produces would escape it; agents reran the reproducers of a random sample,
and \rerunNot{} of \rerunRan{} did not reproduce
(\S\ref{sec:study-scoring}). The sample is small: that neither scorer grades any of its \manualN{} findings false bounds the share of false findings only below 6\%, and that \rerunReproduced{} of its \rerunRunnable{} programs reproduce bounds the share that reproduce only above 80\% (95\% confidence). Every arm has three audits, which differ widely. We run no statistical tests; we state a difference only where it holds for every audit of the arms compared, and otherwise call it a tendency. Each arm adds one section to the bare prompt, so the arms do not isolate interactions between sections; the \emph{Fitness} section of the \emph{+target} arm refers to the triage rules, which that arm does not receive; and the full prompt holds parts that no single arm tests. Scorers do not see the arm, but where an audit put its output (\texttt{.crustify/reviews/} or \texttt{findings/}) can reveal part of it.

\paragraph{Budgets, harness, baseline, and instruments.} The two models ran under different budgets, which neither reached; all six audits of the closed model ended between \$9 and \$11 of their \$15, which suggests that the agent paces itself by the stated budget. The open model may do better in a harness made for it than in Claude Code. We do not compare the agents with a static analyzer or a fuzzer on the same revisions; their targets differ from ours (\S\ref{sec:related}), but a comparison would show what each finds that the others miss. A reproducer shows undefined
behavior only where an instrument checks for it: aliasing violations across the
boundary need BorrowSanitizer, which requires the foreign library to be statically
linked and built with it, and code that no build configuration compiles is not
exercised. Soundness is not exploitability: most of our findings need a
program that misuses the API in a specific way.

%% file: sections/related.tex
\section{Related Work}
\label{sec:related}

\paragraph{LLMs and the soundness of Rust code.} In
deepSURF~\cite{deepsurf2025}, a model augments fuzzing harnesses for the APIs
that reach \texttt{unsafe} code, and the fuzzer finds the bug; its dataset
excludes crates that depend on external software, i.e., the bindings that we
audit. Safe4U~\cite{li2025safe4u} has a model decide, with context retrieved
by static analysis, whether each safety contract of an \texttt{unsafe} call
holds; its output is a verdict, not a program that demonstrates the
defect. Our agents both find and demonstrate the defect.

\paragraph{Agents that audit Rust.} Davidoff~\cite{davidoff2026unreasonable}
audits Rust crates, among them wrappers of C libraries, with a coding agent and
a published prompt, and confirms findings with Miri, the sanitizers, and
Valgrind. Microsoft's FORGE Lab~\cite{microsoft2026forge} reports soundness
bugs in the Rust standard library found with an AI-assisted scanner and
demonstrated in safe Rust, and names the boundary with C and the kernel as a
promising target. BorrowSanitizer's maintainers list bugs that its users found
in FFI crates~\cite{bsanbugs}. Like our Insight~\#1, these reports show that
agents find soundness bugs without a fuzzer or a static analyzer; none
measures what guidance changes, how findings fare against a stated bar, how
models compare, or how reproducible an audit is.

\paragraph{Agentic vulnerability discovery.} Claude Mythos
Preview~\cite{anthropic2026mythos} runs in Claude Code, as our agents do, and
found thousands of high-severity vulnerabilities in operating systems, browsers,
and cryptographic libraries, one of them through \texttt{unsafe} Rust; 90.6\%
of 1,752 assessed reports were valid~\cite{anthropic2026glasswing}. Big
Sleep~\cite{bigsleep2024} found an exploitable bug in SQLite; the AI Cyber
Challenge's systems pair agents with fuzzing~\cite{zhang2026aixccsok};
OSS-Fuzz-Gen~\cite{ossfuzzgen2024} has a model write fuzz targets; and
Revelio~\cite{hou2026revelio} confirms the bugs that its agents propose with
an input that a sanitizer checks. Vendors also offer such scanning to
open-source projects~\cite{anthropic2026ossscanner,mozilla2026zerodays}. All
target bugs that an input to the existing program triggers, not a new program
that misuses an API.

\paragraph{Static detection.} Static analyzers detect the patterns that they
encode, e.g., panic-safety bugs and unsound \texttt{Send} and \texttt{Sync}
bounds, deallocation bugs, type confusion, or memory errors across the FFI
boundary~\cite{bae2021rudra,li2021mirchecker,cui2023safedrop,nitin2024yuga,typepulse2025,li2022ffichecker,unsafechecker2026};
they report candidates rather than demonstrated bugs, and except for
FFIChecker they do not model the foreign library.

\paragraph{Dynamic checking.}
Sanitizers~\cite{serebryany2012asan,serebryany2011tsanllvm,msan2015},
Miri~\cite{miri2026}, and BorrowSanitizer~\cite{mccormack2025bsan}, the
instruments of our agents, report a violation when an execution exhibits it.
McCormack et al.~\cite{mccormack2025mirilli} ran the tests of crates that call
foreign functions under an extended Miri and found 46 instances of undefined
behavior in 37 libraries; our agents instead supply the program that misuses
the API.

\paragraph{Fuzzing Rust libraries.} RULF~\cite{jiang2021rulf},
RPG~\cite{xu2024rpg}, RuMono~\cite{zhang2025rumono}, FRIES~\cite{yin2024fries},
and GRAFT~\cite{chen2026graft} generate fuzz targets for Rust libraries from
their API signatures, generic parameters, or uses in the ecosystem, and find
the bugs that a generated sequence of calls triggers under a sanitizer.

\paragraph{Verification.} RustBelt~\cite{jung2018rustbelt} proves the
soundness of \texttt{unsafe} code in the standard library, and
Kani~\cite{delmas2026kani} model-checks Rust code against developer-written
harnesses; both give guarantees that an audit cannot, at an effort per API that
an audit avoids.

%% file: sections/conclusion.tex
\section{Conclusion}
\label{sec:conclusion}

We studied LLM agents as auditors of Rust code that interfaces with C and C++ and
extended \sys with an auditor built on what the study shows to matter. Agents
already understand the defects of this boundary, and almost nothing that they
file is false; what they lack is direction
and discipline, which named objectives, triage rules, and a stated target
supply. With all of the guidance, the open model finds as much as the closed
model at a quarter of the cost, and a first audit of the kernel's Rust abstractions
suggests that the approach carries over to them.
Audits are only partly reproducible, so several together find more than any
one, and agents that can read a project's tracker restate it. With \sys, we
found \numFindings{} defects in \numCrates{} widely used crates, \numUB{} of
them undefined behavior, and \rflFindings{} in the Rust abstractions of Linux.

%% file: sections/appendix.tex
\section{Ethical Considerations}
\label{sec:ethics}

\paragraph{Stakeholders.} Our audits affect the maintainers of the audited
crates, whose time our reports take, and the users of those crates, who are
exposed while a defect is unfixed and benefit when it is fixed. We audited only
open-source code, ran every reproducer on our own machines, and attacked no
deployed system.

\paragraph{Agents do not report.} An agent with network access could file
issues itself. Ours ran without credentials, so that they could read the
network but not write to it, and before a finding is reported, a person reviews it,
reruns its reproducer, and decides whether and how to report it. Machine-written reports
cost maintainers time when they are wrong, duplicated, or overstated. The
triage rules exist to prevent exactly these failures, and before reporting we
searched each project's tracker, merged findings that share a root cause, and
dropped \numDoNotFile{} confirmed findings that did not merit a report.

\paragraph{Disclosure.}\label{sec:disclosure} We followed each project's security policy, and reported most findings as public issues, one issue per root
cause or per group of findings on one API. Our issues name \sys as their
source, so that maintainers can weigh them accordingly. For three projects (\texttt{curl}, \texttt{rusqlite}, and
\texttt{tree-sitter}), we reported privately first. This paper names only findings that are public or fixed
(Appendix~\ref{sec:reported}); for the \numToFile{} findings that we have not reported yet and for the kernel findings, it gives counts and classes only. Several reported findings are
still unfixed. They are already public in the projects' trackers, and none is
remotely exploitable without an application that misuses the affected API in a
specific way.

\paragraph{Dual use.} The guidance that makes agents better auditors also helps
an attacker find bugs. Soundness bugs, however, are mostly not exploitable on
their own: each needs a program that misuses an API, and the program is under
the control of the crate's user, not of an attacker. We open-source the agent
because the defenders of these crates, their maintainers, gain more from them
than an attacker does.

\input{sections/case-studies-appendix}

\section{Defects by Class in the Study}
\label{sec:classes-study}

Table~\ref{tab:classes-study} counts the distinct real defects of every audit
of the controlled study by class (\S\ref{sec:rq2}).

\begin{table*}[t]
\centering
\caption{Distinct real defects per audit by class. In the arms without the
objectives, the class is the one that the scorer assigned, since these audits
do not use the objectives' names; \emph{panic} and \emph{leak} are kinds that
such audits report on their own. Nits are excluded.}
\label{tab:classes-study}
\small
\begin{tabular}{@{}llcccccc@{}}
\toprule
Model & Guidance & \req{ub} & \req{equiv} & \req{contr.} & Panic & Leak & Other \\
\midrule
\input{data/study-classes}
\end{tabular}
\end{table*}

\paragraph{Memory safety and behavior.} The defects of behavior that
unprompted audits report include errors and panics of callbacks that the
binding swallows, return values that it discards, and option bits that it maps
to the wrong constant. The closed model's own headings are ``memory safety''
and ``logic, error handling, panics''. Audits that are given the objectives
refile panics under \req{equiv}, e.g., 33 of the closed model's full-prompt
findings, and drop most leaks, which accounts for part of the rise of
\req{equiv} in Table~\ref{tab:classes-study}.

\paragraph{Contracts.} Both models find that the contract of the trait method
that builds a wrapper from a raw pointer omits ownership, and each finds one
further contract that the other does not. The scorers rate most contract
findings as nits, which Table~\ref{tab:classes-study} excludes. Of the 17
findings filed under \req{contracts}, \suppContractsArgued{} rest on an
argument alone (\S\ref{sec:rq2}).

\section{Requests, Hazards, and the Judge's Brief}
\label{sec:prompts}

\lstset{basicstyle=\ttfamily\scriptsize, breaklines=true, columns=fullflexible,
  frame=single, framerule=0.3pt}

\paragraph{Request.} The arms with the objectives receive the request below.
The other arms receive the same request with its first sentence ending in
``Audit it for defects.'' The closed model's request states \$15 instead of
\$10.
\begin{lstlisting}
The Rust crate in /target is rust-lang/git2-rs at b9d59b6, a binding to libgit2. Audit it
against these objectives: ub, contracts and equiv. Audit no other objective.
Audit the git2 crate only, the code under src/: libgit2-sys, git2-curl and the vendored
libgit2 are out of scope, except as they explain a defect in git2.

Configuration: --features "https ssh vendored-libgit2 vendored-openssl unstable", and no
other feature.

Put each defect in its own directory under /target/findings/, with a report that could be
sent to the project that owns it, and list them in /target/findings/REPORT.md.
Report every defect that you find, also one that is already reported upstream: if a
report exists, link it.

Budget: $10 and 3 hours of wall-clock time, subagents included. Run `spend` to see
what the run has used so far. Stop when either is spent.
Work without asking me anything. You may read anything on the network, but post nothing:
no issues, comments, pull requests or messages anywhere. The reports stay in /target.
In /target/findings/REPORT.md, also list every suspected defect that you considered and what
became of it.
\end{lstlisting}

\paragraph{Hazards section.} The \emph{+hazards} and \emph{full} arms append
this section, phrased for a binding to a C library.
\begin{lstlisting}
## Hazards of undefined behavior

Rust code that meets C tends to reach undefined behavior through the hazards below. The
list is a starting point: a defect that fits none of them is still a defect.

- **Lifetimes.** An object outlives the owner that it points into. A later call, or a
  second handle to the same C object, invalidates a borrow. C keeps a pointer to a value
  that moves afterwards.
- **Callbacks.** A callback outlives its closure or the data that the closure borrows. A
  callback re-enters the object that C is working on. C calls a closure on another
  thread. A panic unwinds into C.
- **Threads.** An `unsafe impl` of `Send` or `Sync` promises what the C object does not
  have. Two handles share C state without a lock. A safe function touches C globals.
- **Aliasing.** A mutable pointer is derived from a shared borrow. A `&mut` is not
  unique. C writes behind a shared reference.
- **Values from C or from input.** A length, index or offset is not checked. A pointer or
  a function pointer is assumed valid. Data is not a valid value of its Rust type. Memory
  is read before C wrote it.
- **Ownership.** Both sides free one object. A method that is valid once can be called
  twice. Memory is freed with the wrong deallocator. A skipped `Drop` breaks an invariant.
- **Integers.** A conversion truncates or changes sign. An unchecked value makes C
  overflow, divide by zero, or shift too far.
- **Layout.** A Rust type differs from the C type that it is cast to. An access is
  misaligned or too large. An opaque C type is treated as a Rust value.
- **Unsafe APIs.** A caller meets every documented condition and still gets it wrong.
\end{lstlisting}

\paragraph{The judge's brief.} Each judge instance receives this brief, with
the codes of its audits and its output directory filled in; \texttt{<root>}
stands for the directory of the study.
\begin{lstlisting}
You are scoring, by reading only, the outputs of several automated audits of the Rust crate git2-rs (rust-lang/git2-rs at commit b9d59b6, git2 0.21.0, a binding to libgit2). Be sceptical and independent. Do not build or run anything: no cargo, no docker. Post nothing on the network; you may read GitHub issues to check an upstream citation if needed, sparingly.

**Inputs.** Each audit's checkout is a directory `<root>/scoring-f/in/<code>` for the codes: <codes>. An audit's output is in one of two places: under `.crustify/reviews/` (a report `REPORT-<date>.md` and one directory per finding at `<objective>/<slug>/`), or under `findings/` (a `REPORT.md` and one directory per finding). Formats are each audit's own choice. Build directories and scratch were removed from these copies, so a missing binary or cache is not a defect of the audit. Each audit was asked to list in its report every suspected defect it considered and what became of it. Each audit was also asked to report every defect it finds, including ones already reported upstream, and to link the upstream report where one exists; so a finding on a known upstream bug is not an error, but a missing or wrong link is (record it under says_known and known_correct).

Do not read `<root>/runs/`, `<root>/scoring/[A-Z]*` directories or score files, `<root>/scoring-d*` and `scoring-e*` and `scoring-f-key.json` and `scoring-f-groups.json`, other codes' directories, any other `scoring-f/out-g*` directory (do not list or glob `scoring-f/` itself), or this session's task-output directory. Do not try to find out which configuration produced which code.

**Reference table.** `<root>/scoring/adjudicated-root-causes.tsv` (columns id, description, class, gt_slug, gt_visibility, upstream, in_scope2, real, per-audit columns, marked_known, note) with its rules in `adjudication-notes.md` beside it. It lists the root causes known for this revision; it is incomplete (later audits found more), and one row is wrong: `diff-merge-into-parsed-double-free` is upstream issue #964, so treat it as public. Granularity: one row per defect that one coherent fix removes.

**For every audit, and every finding directory in it:** read the finding and the source lines it cites in the checkout, then record: the table `id` of each root cause it covers (several rows if it bundles; a proposed new id with file:line if none fits); the class of the defect (ub / contracts / equiv / panic / leak / other; your reading if the audit names none); the evidence the audit stored, as it stands on disk: `san` (a stored sanitizer or Miri report), `crash` (stored crash or signal), `beh` (stored output showing wrong behaviour, a panic or a leak), `arg` (argument only), `known` (recorded as known upstream, no reproducer); whether a reproducer program is present, whether it contains `unsafe`, whether it declares `#![forbid(unsafe_code)]`; whether the audit says the defect is known upstream and whether that agrees with the table's `upstream` column; and your judgement from reading: real (yes / nit / no / unclear) whether the claimed class or severity is overstated; and whether the finding duplicates, or is split from the same root cause as, another finding in the same audit (name it).

**Then, from each audit's report:** every suspected defect it lists as considered but not filed (refuted, known, out of scope, undecided, dismissed), matched to a table id where one fits, with the audit's reason and whether you agree. Note the real table rows among them.

**Output.** For each code write `<root>/scoring-f/out-g<n>/<code>-scores.tsv` with header `finding	root_cause_id	new_id	class_claimed	evidence	has_repro	uses_unsafe	forbid_unsafe	says_known	known_correct	real	overstated	dup_of	note` (one row per finding and root cause), and `<root>/scoring-f/out-g<n>/<code>-leads.tsv` with header `lead	root_cause_id	outcome_given	reason	agree	real_in_table`. Write one shared `<root>/scoring-f/out-g<n>/notes.md` with anything a reader must know (proposed new ids with file:line, bundles, doubtful matches).

Reply with one compact table, one row per code: findings filed; distinct real root causes; of those, unpublished ground truth (gt_visibility unpublished) and public; new to the table; evidence counts san/crash/beh/arg/known; findings with `forbid_unsafe`; false findings; upstream bugs filed as new; bundles; findings that duplicate or split a root cause already filed in the same audit; leads listed as not filed, and how many of those are real table rows (and how many unpublished). Then at most ten lines of remarks.
\end{lstlisting}

\onecolumn
\section{Reported Findings}
\label{sec:reported}

Table~\ref{tab:reported} lists every finding that we have reported, with the
instrument that fired on its reproducer and its state upstream as of
2026-11-25. \emph{Report} links the issue or pull request; findings that
share one were reported together, and \emph{private} marks findings sent to
the maintainers privately. The table lists the \numReported{} distinct findings that we reported, one row each; a reported finding with the root cause of another row is left out.

{\footnotesize
\begin{longtable}{@{}lp{8cm}llll@{}}
\caption{The findings reported to maintainers.}\label{tab:reported}\\
\toprule
Crate & Finding & Class & Instrument & Report & State \\
\midrule
\endfirsthead
\toprule
Crate & Finding & Class & Instrument & Report & State \\
\midrule
\endhead
\bottomrule
\endfoot
\input{data/reported}
\end{longtable}
}

%% file: sections/case-studies-appendix.tex
\section{Further Case Studies}
\label{sec:more-cases}

\paragraph{A borrow that outlives \texttt{init} (\texttt{openssl}).}
\texttt{X509StoreContextRef::init} takes a trust store, a certificate, and a
chain by shared reference, initializes the context with them, and runs a closure
on it. In OpenSSL, \texttt{X509\_STORE\_CTX\_init} stores raw pointers to all
three in the context, and \texttt{X509\_STORE\_CTX\_cleanup}, which a guard in
\texttt{init} runs on return, clears the chain and the verification parameters
but leaves the certificate pointer set. The rule is that the context may be
used only between the two calls. The binding states the rule in a doc comment
of \texttt{verify\_cert} (``This will only work inside of a call to
\texttt{init}''), but \texttt{verify\_cert} is a safe method that remains
callable after \texttt{init} has returned and the certificate has been dropped
(Listing~\ref{lst:case-openssl}). The bug belongs to the \emph{lifetimes}
hazard: the C side keeps a pointer to a value that Rust frees afterwards.
AddressSanitizer reports a heap-use-after-free WRITE of size 4 in
\texttt{CRYPTO\_UP\_REF}, called from \texttt{X509\_verify\_cert}, on memory
that \texttt{X509\_free} released in the certificate's destructor. The agent
found two further paths to the same root cause: a nested \texttt{init} inside
the closure, which never leaves \texttt{init}, and a call with every argument
still alive, which dereferences the cleared parameters. All three paths also
crash with a segmentation fault against the uninstrumented system
\texttt{libcrypto}, so this bug does not need the instrument to show itself;
AddressSanitizer names the fault as a write to freed memory and shows where it
was freed. The fix, PR~\#2703 in \texttt{openssl}~0.10.82, has the cleanup
guard also clear the certificate and current-certificate pointers, so that
\texttt{verify\_cert} after \texttt{init} returns an error. It closes our two
issues on this root cause, \#2679 and \#2680, without a change to the API.

\begin{figure}[t]
\begin{lstlisting}[language=Rust, label={lst:case-openssl},
  caption={A use-after-free in \texttt{openssl}~0.10.81 (issue \#2680).
  \texttt{self\_signed} builds a certificate in safe code.}]
#![forbid(unsafe_code)]
fn main() -> Result<(), ErrorStack> {
  let mut ctx = X509StoreContext::new()?;
  {
    let cert = self_signed();
    let store = X509StoreBuilder::new()?.build();
    let chain = Stack::<X509>::new()?;
    ctx.init(&store, &cert, &chain, |_| Ok(()))?;
  } // frees cert; ctx->cert still points to it
  let _ = ctx.verify_cert(); // X509_up_ref on it
  Ok(())
}
\end{lstlisting}
\end{figure}

\paragraph{Re-entry during scope teardown (\texttt{mlua}).}
\texttt{Lua::scope} lets Lua call closures that borrow from the caller's stack.
When the scope ends, \texttt{mlua} must invalidate these closures in the Lua
state before it drops the data that they capture, and every constructor on the
scope first checks that the scope is still active, so that no new value
outlives the borrows. The teardown runs in \texttt{Destructors::drop(\&mut
self)}, which takes the list of destructors out of a \texttt{RefCell} and
drops the captured data while \texttt{\&mut self} is live. That data may hold
a \texttt{\&Scope}, and its destructor may call a constructor; upstream
supports this on purpose and tests it. The constructor's check then reads the
\texttt{RefCell} through \texttt{\&Scope}, a foreign read of memory that the
protected \texttt{\&mut} has written (Listing~\ref{lst:case-mlua}). The bug
belongs to the \emph{aliasing} hazard, reached through re-entrancy. The check
returns the correct error; the access that produces it is the defect.
BorrowSanitizer reports that the read access at \texttt{src/scope.rs:250} is
forbidden because it would disable the protected tag created by
\texttt{Destructors::drop}. The crate's own teardown test fails under
BorrowSanitizer in the same way. No other instrument observes the bug: plain
and AddressSanitizer builds run to completion, since no memory is corrupted,
and Miri cannot run the program because it executes Lua's C code. The fix,
commit~\texttt{e4699f2}, which closes our issue \#765, changes two lines: the
list moves behind a \texttt{Box}, out of the memory that the protected
reference covers, as our report proposed.

\begin{figure}[t]
\begin{lstlisting}[language=Rust, label={lst:case-mlua},
  caption={A Tree Borrows violation in \texttt{mlua}~0.12.2 (issue \#765).}]
#![forbid(unsafe_code)]
struct OnDrop<F: FnOnce()>(Option<F>);
impl<F: FnOnce()> Drop for OnDrop<F> {
  fn drop(&mut self) { (self.0.take().unwrap())(); }
}
fn main() -> mlua::Result<()> {
  let lua = mlua::Lua::new();
  lua.scope(|scope| {
    let guard = OnDrop(Some(move || { // at teardown
      let f = scope.create_function(|_, ()| Ok(()));
      assert!(f.is_err());
    }));
    scope.create_function(move |_, ()| { let _ = &guard; Ok(()) })?;
    Ok(())
  })
}
\end{lstlisting}
\end{figure}

%% file: data/study-classes.tex
Closed & none & 15 / 21 / 14 & 2 / 7 / 8 & 0 / 0 / 0 & 8 / 12 / 16 & 4 / 3 / 3 & 6 / 1 / 3 \\
Closed & full & 14 / 17 / 15 & 13 / 18 / 17 & 0 / 1 / 0 & 0 / 0 / 0 & 0 / 0 / 0 & 0 / 0 / 0 \\
Open & none & 18 / 13 / 23 & 4 / 3 / 8 & 0 / 0 / 0 & 11 / 1 / 6 & 5 / 3 / 3 & 4 / 0 / 0 \\
Open & +objectives & 19 / 13 / 16 & 5 / 2 / 2 & 0 / 0 / 0 & 0 / 0 / 0 & 0 / 0 / 0 & 0 / 0 / 0 \\
Open & +hazards & 19 / 20 / 21 & 8 / 4 / 4 & 0 / 0 / 0 & 10 / 13 / 11 & 3 / 4 / 4 & 0 / 0 / 0 \\
Open & +triage & 12 / 14 / 18 & 4 / 3 / 5 & 0 / 0 / 0 & 0 / 9 / 4 & 0 / 3 / 3 & 0 / 0 / 0 \\
Open & +target & 23 / 25 / 22 & 11 / 9 / 11 & 0 / 0 / 0 & 16 / 15 / 11 & 4 / 5 / 5 & 4 / 1 / 2 \\
Open & full & 18 / 26 / 25 & 11 / 19 / 15 & 0 / 0 / 0 & 0 / 0 / 0 & 0 / 0 / 0 & 6 / 0 / 0 \\
\bottomrule

%% file: data/reported.tex
\texttt{aws-lc-rs} & \nolinkurl{aes-key-assume-init-uninit-tail} & UB (safe) & Miri & \href{https://github.com/aws/aws-lc-rs/issues/1326}{\#1326} & open \\
\texttt{aws-lc-rs} & \nolinkurl{des-weak-key-parity-bypass} & Equivalence & assertion & \href{https://github.com/aws/aws-lc-rs/issues/1300}{\#1300} & open \\
\texttt{aws-lc-rs} & \nolinkurl{rsa-public-key-components-skip-key-check} & Equivalence & assertion & \href{https://github.com/aws/aws-lc-rs/issues/1299}{\#1299} & open \\
\texttt{curl} & \nolinkurl{easy2-get-ref-uaf-across-perform} & UB (safe) & ASan/UBSan & \href{https://github.com/sagebind/curl-rust/issues/697}{\#697} & open \\
\texttt{curl} & \nolinkurl{easy2-handler-reentrancy-uaf} & UB (safe) & ASan/UBSan & \href{https://github.com/sagebind/curl-rust/issues/698}{\#698} & open \\
\texttt{curl} & \nolinkurl{getinfo-str-uaf-after-perform} & UB (safe) & ASan/UBSan & \href{https://github.com/sagebind/curl-rust/issues/658#issuecomment-7067369230}{\#658} & open \\
\texttt{curl} & \nolinkurl{message-easy-handle-uaf} & UB (safe) & ASan/UBSan & \href{https://github.com/sagebind/curl-rust/issues/696}{\#696} & open \\
\texttt{curl} & \nolinkurl{multi-detachguard-drop-in-callback-uaf} & UB (safe) & ASan/UBSan & \href{https://github.com/sagebind/curl-rust/issues/699}{\#699} & open \\
\texttt{ffmpeg-next} & \nolinkurl{dictionary-derefmut-borrowed-into-owned} & UB (safe) & Valgrind & \href{https://github.com/zmwangx/rust-ffmpeg/issues/282}{\#282} & open \\
\texttt{ffmpeg-next} & \nolinkurl{filter-context-unbounded-lifetime-uaf} & UB (safe) & Valgrind & \href{https://github.com/zmwangx/rust-ffmpeg/issues/281}{\#281} & open \\
\texttt{ffmpeg-next} & \nolinkurl{frame-setters-desync-slice-bounds} & UB (safe) & ASan/UBSan & \href{https://github.com/zmwangx/rust-ffmpeg/issues/283}{\#283} & open \\
\texttt{ffmpeg-next} & \nolinkurl{metadata-invalid-utf8-str} & UB (safe) & other & \href{https://github.com/zmwangx/rust-ffmpeg/issues/286}{\#286} & open \\
\texttt{ffmpeg-next} & \nolinkurl{packetiter-input-context-layout-confusion} & UB (safe) & other & \href{https://github.com/zmwangx/rust-ffmpeg/issues/287}{\#287} & open \\
\texttt{ffmpeg-next} & \nolinkurl{sample-buffer-plane-index-oob} & UB (safe) & ASan/UBSan & \href{https://github.com/zmwangx/rust-ffmpeg/issues/284}{\#284} & open \\
\texttt{ffmpeg-next} & \nolinkurl{stream-chapter-mut-aliasing} & UB (safe) & Miri & \href{https://github.com/shssoichiro/ffmpeg-the-third/issues/158}{\#158} & open \\
\texttt{ffmpeg-next} & \nolinkurl{subtitle-rect-null-text-deref} & UB (safe) & Valgrind & \href{https://github.com/zmwangx/rust-ffmpeg/issues/285}{\#285} & open \\
\texttt{git2} & \nolinkurl{config-outlives-repository-uaf} & UB (safe) & ASan/UBSan & \href{https://github.com/rust-lang/git2-rs/issues/1331}{\#1331} & open \\
\texttt{git2} & \nolinkurl{diff-send-repository-race} & UB (safe) & TSan & \href{https://github.com/rust-lang/git2-rs/issues/1333}{\#1333} & fixed \\
\texttt{git2} & \nolinkurl{fetchhead-foreach-unnamed-entry-swallowed} & Equivalence & assertion & \href{https://github.com/rust-lang/git2-rs/issues/1340}{\#1340} & open \\
\texttt{git2} & \nolinkurl{odb-sync-mempack-race} & UB (safe) & TSan & \href{https://github.com/rust-lang/git2-rs/issues/1338}{\#1338} & open \\
\texttt{git2} & \nolinkurl{odb-writer-finalize-twice-null-call} & UB (safe) & ASan/UBSan & \href{https://github.com/rust-lang/git2-rs/issues/1337}{\#1337} & open \\
\texttt{git2} & \nolinkurl{rebase-operation-uaf} & UB (safe) & ASan/UBSan & \href{https://github.com/rust-lang/git2-rs/issues/1334}{\#1334} & open \\
\texttt{git2} & \nolinkurl{rebase-options-outlive-uaf} & UB (safe) & ASan/UBSan & \href{https://github.com/rust-lang/git2-rs/issues/1335}{\#1335} & open \\
\texttt{git2} & \nolinkurl{refdb-shared-across-repositories-use-after-free} & UB (safe) & ASan/UBSan & \href{https://github.com/rust-lang/git2-rs/issues/1336}{\#1336} & open \\
\texttt{git2} & \nolinkurl{repo-get-attr-uaf} & UB (safe) & ASan/UBSan & \href{https://github.com/rust-lang/git2-rs/issues/1330}{\#1330} & open \\
\texttt{git2} & \nolinkurl{repo-set-namespace-uaf} & UB (safe) & ASan/UBSan & \href{https://github.com/rust-lang/git2-rs/issues/1330}{\#1330} & open \\
\texttt{git2} & \nolinkurl{repo-set-workdir-uaf} & UB (safe) & ASan/UBSan & \href{https://github.com/rust-lang/git2-rs/issues/1330}{\#1330} & open \\
\texttt{git2} & \nolinkurl{repository-send-shares-index-across-threads} & UB (safe) & TSan & \href{https://github.com/rust-lang/git2-rs/issues/1332}{\#1332} & open \\
\texttt{git2} & \nolinkurl{transport-factory-error-dropped} & Equivalence & assertion & \href{https://github.com/rust-lang/git2-rs/issues/1341}{\#1341} & fixed \\
\texttt{mlua} & \nolinkurl{callback-finalizer-resurrection-uaf} & UB (safe) & ASan/UBSan & \href{https://github.com/mlua-rs/mlua/issues/755}{\#755} & fixed \\
\texttt{mlua} & \nolinkurl{destroy-during-scoped-borrow-uaf} & UB (safe) & ASan/UBSan & \href{https://github.com/mlua-rs/mlua/issues/756}{\#756} & fixed \\
\texttt{mlua} & \nolinkurl{exec-raw-panic-unwinds-through-lua} & UB (unsafe) & ASan/UBSan & \href{https://github.com/mlua-rs/mlua/issues/762}{\#762} & fixed \\
\texttt{mlua} & \nolinkurl{lua-clone-during-close-reenters-drop} & UB (safe) & ASan/UBSan & \href{https://github.com/mlua-rs/mlua/issues/759}{\#759} & fixed \\
\texttt{mlua} & \nolinkurl{main-thread-resumed-as-coroutine} & UB (safe) & other & \href{https://github.com/mlua-rs/mlua/issues/761}{\#761} & fixed \\
\texttt{mlua} & \nolinkurl{outer-guard-released-before-inner-guard} & UB (safe) & TSan & \href{https://github.com/mlua-rs/mlua/issues/763}{\#763} & fixed \\
\texttt{mlua} & \nolinkurl{resume-stale-nargs-oob} & UB (safe) & ASan/UBSan & \href{https://github.com/mlua-rs/mlua/issues/758}{\#758} & fixed \\
\texttt{mlua} & \nolinkurl{scope-teardown-reentry-aliasing} & UB (safe) & BSan & \href{https://github.com/mlua-rs/mlua/issues/765}{\#765} & fixed \\
\texttt{mlua} & \nolinkurl{userdata-metatable-ref-freed-by-finalizer} & UB (safe) & ASan/UBSan & \href{https://github.com/mlua-rs/mlua/issues/764}{\#764} & fixed \\
\texttt{mlua} & \nolinkurl{userdata-ref-guard-protected-across-free} & UB (safe) & BSan & \href{https://github.com/mlua-rs/mlua/issues/760}{\#760} & fixed \\
\texttt{nix} & \nolinkurl{cmsg-buffer-misaligned-cmsghdr} & UB (safe) & ASan/UBSan & \href{https://github.com/nix-rust/nix/issues/2875}{\#2875} & open \\
\texttt{nix} & \nolinkurl{getsockopt-ffi-ptr-invalidated} & UB (safe) & Miri & \href{https://github.com/nix-rust/nix/issues/2878}{\#2878} & open \\
\texttt{nix} & \nolinkurl{select-nfds-overflows-fdset} & UB (safe) & other & \href{https://github.com/nix-rust/nix/issues/2879}{\#2879} & open \\
\texttt{nix} & \nolinkurl{sendmmsg-recvmmsg-dangling-msg-name} & UB (safe) & ASan/UBSan & \href{https://github.com/nix-rust/nix/issues/2871}{\#2871} & open \\
\texttt{nix} & \nolinkurl{sockaddrstorage-from-raw-none-uninit-sun-len} & UB (unsafe) & Miri & \href{https://github.com/nix-rust/nix/pull/2869#issuecomment-7056263503}{\#2869} & open \\
\texttt{nix} & \nolinkurl{sockopt-impl-unchecked-value-type} & UB (safe) & Miri & \href{https://github.com/nix-rust/nix/issues/2877}{\#2877} & open \\
\texttt{nix} & \nolinkurl{unixaddr-from-raw-len-overflow} & UB (safe) & ASan/UBSan & \href{https://github.com/nix-rust/nix/issues/2876}{\#2876} & open \\
\texttt{openssl} & \nolinkurl{aad-update-null-output-pointer} & UB (safe) & ASan/UBSan & \href{https://github.com/rust-openssl/rust-openssl/issues/2670}{\#2670} & open \\
\texttt{openssl} & \nolinkurl{aes-wrap-key-shared-reborrow-write} & UB (safe) & BSan & \href{https://github.com/rust-openssl/rust-openssl/issues/2672}{\#2672} & open \\
\texttt{openssl} & \nolinkurl{dsa-public-key-null-params} & UB (safe) & ASan/UBSan & \href{https://github.com/rust-openssl/rust-openssl/issues/2681}{\#2681} & fixed \\
\texttt{openssl} & \nolinkurl{libctx-lifetime-use-after-free} & UB (safe) & ASan/UBSan & \href{https://github.com/rust-openssl/rust-openssl/issues/2678}{\#2678} & open \\
\texttt{openssl} & \nolinkurl{pkey-ctx-verify-recover-output-overflow} & UB (safe) & ASan/UBSan & \href{https://github.com/rust-openssl/rust-openssl/issues/2671}{\#2671} & fixed \\
\texttt{openssl} & \nolinkurl{read-early-data-shared-reborrow-write} & UB (safe) & BSan & \href{https://github.com/rust-openssl/rust-openssl/issues/2672}{\#2672} & open \\
\texttt{openssl} & \nolinkurl{wrap-mode-update-size-underestimates-des3-wrap} & UB (safe) & ASan/UBSan & \href{https://github.com/rust-openssl/rust-openssl/issues/2682}{\#2682} & fixed \\
\texttt{openssl} & \nolinkurl{x509-store-context-verify-cert-use-after-free} & UB (safe) & ASan/UBSan & \href{https://github.com/rust-openssl/rust-openssl/issues/2680}{\#2680} & fixed \\
\texttt{rocksdb} & \nolinkurl{transaction-iterator-shared-writes} & UB (safe) & ASan/UBSan & \href{https://github.com/rust-rocksdb/rust-rocksdb/issues/1122}{\#1122} & open \\
\texttt{rusqlite} & \nolinkurl{column-metadata-dangling-cstr} & UB (safe) & ASan/UBSan & \href{https://github.com/rusqlite/rusqlite/issues/1950}{\#1950} & fixed \\
\texttt{rusqlite} & \nolinkurl{functions-aggregate-missing-send} & UB (safe) & TSan & private & fixed \\
\texttt{rusqlite} & \nolinkurl{invalues-dangling-valueref} & UB (unsafe) & ASan/UBSan & \href{https://github.com/rusqlite/rusqlite/issues/1951}{\#1951} & fixed \\
\texttt{rusqlite} & \nolinkurl{null-statement-clear-bindings} & UB (safe) & ASan/UBSan & \href{https://github.com/rusqlite/rusqlite/issues/1955}{\#1955} & fixed \\
\texttt{rusqlite} & \nolinkurl{session-start-strm-dangling-reader} & UB (safe) & ASan/UBSan & \href{https://github.com/rusqlite/rusqlite/issues/1954}{\#1954} & fixed \\
\texttt{rusqlite} & \nolinkurl{trace-busy-clientdata-reentrant-uaf} & UB (safe) & ASan/UBSan & private & fixed \\
\texttt{socket2} & \nolinkurl{control-len-counts-unwritten-padding} & Contracts & other & \href{https://github.com/rust-lang/socket2/issues/695}{\#695} & open \\
\texttt{socket2} & \nolinkurl{sockaddr-eq-hash-read-padding} & UB (unsafe) & Miri & \href{https://github.com/rust-lang/socket2/issues/693}{\#693} & open \\
\texttt{socket2} & \nolinkurl{unix-path-bytes-past-sun-path} & UB (unsafe) & Miri & \href{https://github.com/rust-lang/socket2/issues/694}{\#694} & open \\
\texttt{ssh2} & \nolinkurl{error-raw-session-pointer} & UB (safe) & ASan/UBSan & \href{https://github.com/rust-lang/ssh2-rs/issues/380}{\#380} & open \\
\texttt{ssh2} & \nolinkurl{init-atexit-exit-races-session-new} & UB (safe) & TSan & \href{https://github.com/rust-lang/ssh2-rs/issues/383}{\#383} & open \\
\texttt{ssh2} & \nolinkurl{kbdint-empty-prompt-null-slice} & UB (safe) & other & \href{https://github.com/rust-lang/ssh2-rs/issues/385}{\#385} & open \\
\texttt{ssh2} & \nolinkurl{knownhosts-check-without-session-lock} & UB (safe) & TSan & \href{https://github.com/rust-lang/ssh2-rs/issues/382}{\#382} & open \\
\texttt{ssh2} & \nolinkurl{opt-bytes-anchor-outlives-c-string} & UB (safe) & ASan/UBSan & \href{https://github.com/rust-lang/ssh2-rs/issues/378}{\#378} & open \\
\texttt{ssh2} & \nolinkurl{session-send-non-send-stream} & UB (safe) & TSan & \href{https://github.com/rust-lang/ssh2-rs/issues/381}{\#381} & open \\
\texttt{ssh2} & \nolinkurl{sftp-file-handle-readdir} & UB (safe) & ASan/UBSan & \href{https://github.com/rust-lang/ssh2-rs/issues/379}{\#379} & open \\
\texttt{v8} & \nolinkurl{regexp-exec-null-as-object} & UB (safe) & other & \href{https://github.com/denoland/rusty_v8/issues/2103}{\#2103} & open \\
\texttt{v8} & \nolinkurl{resizable-backing-store-shrink} & UB (safe) & other & \href{https://github.com/denoland/rusty_v8/issues/2105}{\#2105} & open \\
\texttt{v8} & \nolinkurl{value-deserializer-data-lifetime} & UB (safe) & other & \href{https://github.com/denoland/rusty_v8/issues/2104}{\#2104} & open \\
\texttt{v8} & \nolinkurl{valueview-gc-use-after-free} & UB (safe) & other & \href{https://github.com/denoland/rusty_v8/issues/2101}{\#2101} & open \\
\texttt{v8} & \nolinkurl{write-v2-null-terminate-overflow} & UB (safe) & other & \href{https://github.com/denoland/rusty_v8/issues/2102}{\#2102} & open \\
\texttt{zstd} & \nolinkurl{block-api-after-one-shot-uaf} & UB (unsafe) & ASan/UBSan & \href{https://github.com/gyscos/zstd-rs/issues/415}{\#415} & fixed \\
\texttt{zstd} & \nolinkurl{noop-run-capacity-reread-overflow} & UB (unsafe) & Miri & \href{https://github.com/gyscos/zstd-rs/issues/416}{\#416} & fixed \\